\documentclass[a4paper,9pt]{article}
\usepackage[a4paper, total={6in, 9.5in}]{geometry}
\usepackage{tabularx}
\usepackage[english]{babel}
\usepackage[utf8]{inputenc}
\usepackage{graphicx}
\usepackage{listings}
\usepackage{verbatim}
\usepackage{amsmath,amsfonts,amssymb}
\usepackage{epstopdf}
\usepackage{placeins}
\usepackage{url}
\usepackage{hyperref}
\usepackage{float}
\usepackage{titling}
\usepackage{amsmath}
\usepackage{multirow}
\usepackage{booktabs}
\usepackage{xcolor}
\usepackage{tikz}
\usepackage[affil-it]{authblk}
\usepackage[font={footnotesize}]{caption}
\usepackage{makecell}
\usepackage{array}
\usepackage{cite}
\usepackage{multirow}
\usepackage{arydshln}

\renewcommand{\vec}[1]{\underline{#1}}
\definecolor{modulecol}{RGB}{200,200,200}
\definecolor{embedcol}{RGB}{190,120,190}
\definecolor{inputcol}{RGB}{250,180,250}
\definecolor{outputcol}{RGB}{150,150,255}
\newcolumntype{Y}{>{\raggedright\arraybackslash}X} 

\newcommand{\C}{\text{C}}
\newcommand{\B}{\text{B}}

\title{Learning relationships in epidemiological data using graph neural networks}
\author[1]{A. J. Wood}
\author[1]{A. R. Sanchez}
\author[1,2]{R. R. Kao\footnote{Corresponding author: rowland.kao@ed.ac.uk}}
\affil[1]{Roslin Institute, University of Edinburgh}
\affil[2]{School of Physics and Astronomy, University of Edinburgh}

\date{\today}

\begin{document}
	
	\maketitle
	
	\begin{abstract}
		\begin{itemize}
		When designing control strategies for an infectious disease it is critical to identify the key pathways of transmission. Data on infected hosts -- when they were born, where they lived and with whom they interacted -- can help infer sources of infection and transmission clusters. However such data are generally not powerful enough to identify infector-infectee pairs with any certainty. 
		
		Whole-genome sequencing data of the underlying pathogen, on the other hand, can serve as a powerful adjoint to these data as they can be used to estimate a time to a most recent common ancestor between two infected hosts. and in turn their relative proximity in the transmission tree. A statistical model that explains the genetic distance between different host pathogens and associated risk factors can therefore inform key risk factors for transmission itself.
		
		We show how graph neural networks (GNNs) are a powerful and natural modelling architecture for such a problem. By treating the epidemiological dataset as a graph where infected hosts are nodes and edges are weighted by the genetic distance between different host pairs, we show how a GNN can be fit to predict the genetic distance between known hosts and new, unsequenced hosts. Comparisons with other established approaches show that GNNs have useful performance advantages albeit with greater computational cost.
		\end{itemize}
	\end{abstract}
	
\section{Introduction}
\subsubsection*{Utility of whole-genome sequencing in precision epidemiology}
Dense whole-genome sequencing (WGS) of infectious disease pathogens is now possible on a mass-scale, with, as of mid-2025, over 15 million SARS-CoV2 sequences shared on GISAID alone. Exploiting the strong topological relationships between pathogen evolutionary trees and infector-infectee transmission trees, WGS is now an established, powerful tool for precision epidemiology. For example, even when epidemiological characteristics such as the generation time, incubation period and serial intervals are well established, the available life history data on a host may be consistent with multiple plausible infectors, making exact `who-infected-whom' inference difficult. In this case the genetic distances between their pathogens can then be used to dramatically reduce the possible sources of infection~\cite{wood2024utility}.

One of the main methodological challenges in precision epidemiology is that the disease datasets of interest are rarely complete; in a given outbreak many infected hosts are likely unidentified, and of those that were identified, not all may have quality metadata. In these settings inference to characterise the full transmission tree is unfeasible as many events will have no data associated with them at all. The evolutionary phylogenetic tree of host pathogens however can serve as an approximation for the structure of the transmission tree. Even without knowing the full transmission tree, then, there is value in understanding patterns in genetic variation in disease data, and statistical models developed for this purpose offer two key insights. First, they can identify risk factors associated with genetic similarity, and by proxy they also identify potential transmission pathways. Second, for a new infected host from which no pathogen sequence data are available, such models could be used to identify the most likely position of the unobserved sequence on he transmission tree and the uncertainty in that position, under the assumption that the statistical model is relevant for the sample ~\cite{rossi2022phylodynamic, rossi2023unraveling, warren2023spatial, gamza2024infector, gamza2024using, degruttola2023modeling, rich2023network, leavitt2022can,steinberg2025regression}.

\subsubsection*{Modelling the genetic relationship between host pairs}
The genetic distance between two sampled pathogens (i.e. the number of evolutionary changes that separate them on a phylogenetic tree) is the sum of the total branch lengths connecting them in the tree. Only some of these relate to direct transmission between hosts or relatively short transmission chains, and therefore some caution must be applied when attempting to infer `who-infected-whom' relationships. However, previous work has shown that they do provide valuable insights into transmission patterns as they are related to explanatory variables linked to transmission risks. Having $H$ hosts with WGS samples and sufficient metadata, a conventional way to organise the data for training a statistical model is as $\frac{1}{2}H\times(H-1)$ rows, with each row associated to a different pair of hosts (hosts A and B, hosts A and C, hosts C and B etc.), with the genetic distance between their pathogens, and pairwise relations (e.g.~physical distance, mean sample time, difference in sample times, duration of interaction). This is a natural structure for a suite of standard modelling approaches e.g.~boosted regression trees, random forests, where a collection of variables $(x_1, x_2, x_3,\dots)$ are used to explain variation in something else $(y)$. This representation can be extended to associate relative branch lengths to the relation, by considering the number of substitutions observed in host A sample that are not seen in B, relative to a reference sequence~\cite{wood2024utility}.

While amenable for model training, from a statistical modelling perspective this pairwise framework treats each host-host relationship as an independent observation. However, for an infectious disease the underlying system from which these data have been drawn is one where all hosts are intrinsically connected in a single tree-like structure. By effectively treating the data as a set of $\frac{1}{2}H\times(H-1)$ observations, then, we do not use any potential contextual information provided from other hosts in the dataset when making a prediction on a host pair's relationship.  As a simple illustrative example consider three hosts A, B, C. Suppose A and B's pathogens are closely genetically related, but A and C are genetically distinct. By using host A for context it follows logically that B and C are also likely to be distinct, irrespective of other host data. In a pairwise model, these relationships may have been misclassified if relations A--B, B--C and C--A were treated as independent.

\subsubsection*{Aim of this work}
In this study we explore statistical modelling approaches that would fit genetic variation between host pairs while preserving the full relational structure of the dataset. Our specific interest is in the potential application of graph neural networks (GNNs)~\cite{scarselli2008graph, wu2020comprehensive}, which to our knowledge are hitherto unexplored in this context. GNNs are a sub-class of neural networks that accept graph-structured data (i.e.~a set of nodes, and relations between those nodes) as inputs. They can be deployed to predict properties of the graph as a whole~\cite{wieder2020compact, tang2023application, zhang2023ss}, properties of specific nodes~\cite{zhang2023graph, xiao2022graph}, or of the edges (relationships between nodes)~\cite{huang2024link, wen2021bondnet}. In this context, nodes represent infected hosts and the genetic distance between two hosts' pathogens is an edge property, making this an edge-level task.

We use infectious disease datasets on bovine Tuberculosis (bTB) in Great Britain. bTB is endemic in domesticated cattle and wild badgers in parts of Great Britain~\cite{broughan2016review} and is an especially complex disease for infector-infectee inference, with complicating factors including substantial between-species transmission~\cite{van2021inferring} and poor sensitivity of the standard test for disease in cattle~\cite{de2006ante}. Furthermore, the causative bacterial pathogen (\emph{Mycobacterium bovis}) exhibits slow and highly variable evolution with only one variation on the genome every $\sim$1--10 years~\cite{crispell2017using, crispell2019combining, trewby2016use, akhmetova2023genomic, canini2023deciphering}.

This paper is structured as follows. In Section~\ref{sec:Data} we describe the mix of synthetic and real-world bTB datasets we fit our models to. In Section~\ref{sec:Methods} we outline a graph neural network approach to fitting pathogen sequence data, adapting an existing neural network architecture to our epidemiological modelling task. We present the methodology in a way that is accessible to working epidemiologists without requiring prior technical experience in neural networks, and without being tied to any specific disease context. In Section~\ref{sec:Results} we compare the predictive power of GNNs to a set of pairwise statistical models (logistic regression, random forest, boosted regression tree), and use permutation importance to estimate the explanatory power of different variables in the dataset across different models. We find that, while GNNs are able to make more accurate predictions on host-host relations in the exemplars we tested, this declines substantially for smaller datasets. In Section~\ref{sec:Discussion} we discuss how the permutation importance indicates that for these larger datasets the GNN is leveraging additional context from the broader disease dataset to make more accurate predictions on host-host relations. Finally we discuss limitations to this modelling approach, and other potential uses of the GNN architecture in the context of infectious disease data.

\section{Data}\label{sec:Data}
The datasets used in this work are summarised in Table~\ref{tab:data}, with associated plots and graph representations of genetic variation provided in Supplementary Material, Section~\ref{supp:Data},~Figures~\ref{fig:dataS1},~\ref{fig:dataS2},~\ref{fig:dataS3},~\ref{fig:dataW},~\ref{fig:dataC}. These comprise three synthetic datasets and two real-world datasets on bovine Tuberculosis (bTB) disease. Note that the number of edges grows as the number of hosts squared. The explanatory variables associated with each dataset are provided in Supplementary Material, Tables~\ref{tab:attributes_synthetic},~\ref{tab:attributes_woodchester},~\ref{tab:attributes_cumbria}.

Each of the synthetic datasets comprise $H = 2\,000$ hosts (being the first $2\,000$ hosts sampled in each simulation) with perfect coverage. The two real-world datasets are substantially smaller ($H=241$ and $63$ hosts respectively). For the Woodchester dataset, owing to the much larger genetic diversity of samples collected (median genetic distance of 26 SNPs), a close host pair is defined as one with a genetic distance of 4 SNPs or fewer.

The Woodchester Park data can be thought of as being drawn from an open system where infection can come from a host within the study area, but also from outside the area, as it is embedded within a much larger area where bTB is endemic in both cattle and badgers. Conversely the Cumbria samples were gathered from a novel bTB outbreak in an area with no prior recent history of infection, and so it can be considered a closed system~\cite{rossi2022phylodynamic}.

\begin{table}[t]
    \centering
    {
    \sffamily
	\footnotesize
	\begin{tabularx}{\textwidth}{|p{2.7cm}|p{3.5cm}|Y|}
		\hline
		\textbf{Dataset} & \textbf{Size} & \textbf{Description} \\
		\hline
		Synthetic 1 (badger-driven dynamics)\vspace{0.5cm} & $H$ = 2\,000 (1\,537 cows, 463 badgers, 1\,999\,000 host pairs) & \multirow{3}{=}{Figures~\ref{fig:dataS1},~\ref{fig:dataS2},~\ref{fig:dataS3}. Datasets generated by the Tuberculosis Modelling Initiative simulation model (TBMI, technical details in Supplementary Material, Section~\ref{sec:TBMI}). This model permits transmission between cattle and badgers, as well as within each species. Cattle populations and movements mirror those in GB from 2006--2020. Pathogen evolution is modelled as a Poisson process, at rate of 0.3 SNPs per genome per year. Outbreaks are simulated in the Cumbria area.} \\
		\cline{1-2}
		Synthetic 2 (cattle-driven dynamics)\vspace{0.5cm} & $H$ = 2\,000 (628 cows, 1\,372 badgers, 1\,999\,000 host pairs) & \\
		\cline{1-2}
		Synthetic 3 (hybrid-driven dynamics)\vspace{0.5cm} & $H$ = 2\,000 (1\,409 cows, 591 badgers, 1\,999\,000 host pairs) & \\
		\hline
		Woodchester & $H$ = 241 (130 cows, 111 badgers, 28\,920 host pairs) & Figure~\ref{fig:dataW}. Samples taken between 2000 and 2020 in and around Woodchester Park, south-west England, within a region where bTB is endemic in cattle and badgers. This has been the centre of a long-term capture-test-release study of badgers aimed at better understanding the dynamics of \emph{M. bovis} transmission. Median genetic distance between host pairs is 26 SNPs. 783 host pairs have a genetic distance of 4 SNPs or fewer (class ratio 1:36.0). This is the same dataset used by the authors in Reference~\cite{wood2024utility} where more details can be found. \\
		\hline
		Cumbria & $H$ = 63 (24 cows, 39 badgers, 1\,953 host pairs) & Figure~\ref{fig:dataC}. Samples taken between 2014 and 2019 in Cumbria, north-west England. This was a novel outbreak of bTB in both badgers and cattle, in an area with no recent history of infection. The median genetic distance between host pairs is 2 SNPs. 232 host pairs have a genetic distance of 0 SNPs (class ratio 1:7.4). \\
		\hline
	\end{tabularx}
    }
	\caption{Summary of the datasets used in this work.}
	\label{tab:data}
\end{table}

\section{Methods}\label{sec:Methods}
We classify host pairs as closely related based on the genetic distance between their \emph{M. bovis} WGS samples, in terms of the single-nucleotide polymorphism (SNP) distance between them (the threshold varying depending on the data used). As closely related host pairs comprise a small proportion of all pairings this classification problem is imbalanced, with negatives heavily outweighing positives.

For brevity we say the genetic distance between two hosts is the genetic distance between those hosts' pathogen WGS samples.

\subsection{Graph neural network model for epidemiological data}
Neural networks are a machine learning framework where input data are passed through a series of linear (e.g.~addition, averaging, concatenation, matrix multiplication) and non-linear (e.g.~$x \mapsto \text{sigmoid}(x)$, $x\to \text{max}(x,0)$) transformations~\cite{Goodfellow-et-al-2016}. At each step the dimensionality of the data typically changes. The structure of the final output depends on the task, but for binary classification it is a scalar value between 0 and 1. Generally speaking the intermediate outputs of the network are not interpretable. Graph neural networks are a specialisation of neural networks designed to admit data with a graph structure, and data transformations explicitly account for the relationships between different nodes and edges.

\subsubsection*{Organising an epidemiological dataset into a graph}
Take an epidemiological dataset with $H$ samples labelled $(h_1, h_2, \dots h_H)$, where each sample corresponds to an infected host. Each host has a pathogen WGS sample. With $H$ hosts there are $\frac{1}{2}H(H-1)$ unique undirected pairings of those hosts $(i,j)$ (twice as many if pairings are directed). The genetic distance $d_{ij}$ between each of the host pairs is naturally written as a matrix $\mathbf{D}$:
\begin{align*}
	\mathbf{D} = 
	\begin{bmatrix}
		\cdot    & \cdot    & \cdot    & \cdot    & \cdot    & \cdots & \cdot & \cdot \\
		d_{21}   & \cdot    & \cdot    & \cdot    & \cdot    & \cdots & \cdot & \cdot \\
		d_{31}   & d_{32}   & \cdot    & \cdot    & \cdot    & \cdots & \cdot & \cdot \\
		d_{41}   & d_{42}   & d_{43}   & \cdot    & \cdot    & \cdots & \cdot & \cdot \\
		d_{51}   & d_{52}   & d_{53}   & d_{54}   & \cdot    & \cdots & \cdot & \cdot \\
		\vdots   & \vdots   & \vdots   & \vdots   & \vdots   & \ddots & \vdots & \vdots \\
		d_{H1}   & d_{H2}   & d_{H3}   & d_{H4}   & d_{H5}   & \cdots & d_{H(H-1)} & \cdot
	\end{bmatrix}
\end{align*}
populating the lower diagonal only. Each individual host $i$ has a set of $P$ attributes $\vec{n}_i = (n_{i,1},n_{i,2},n_{i,3},\dots n_{i,P})$, denoted with a tensor $\mathbf{N}$:
\begin{align*}
	\mathbf{N} = 
	\begin{bmatrix}
	\vec{n}_1 \\
	\vec{n}_2 \\
	\vec{n}_3 \\
	\vec{n}_4 \\
	\vdots    \\
	\vec{n}_H
	\end{bmatrix}\;.
\end{align*}
Example node attributes may be the time the host was sampled, the host's $x$ and $y$ positions at that time, and species. Then, each host pair $(i,j)$ has a set of $Q$ relational attributes $\vec{e}_{ij} = (e_{ij,1}, e_{ij,2}, e_{ij,3}, \dots e_{ij,Q})$, denoted with a tensor $\mathbf{E}$:
\begin{align*}
	\mathbf{E} = 
	\begin{bmatrix}
		\cdot    & \cdot    & \cdot    & \cdot    & \cdot    & \cdots & \cdot & \cdot \\
		\vec{e}_{21}   & \cdot    & \cdot    & \cdot    & \cdot    & \cdots & \cdot & \cdot \\
		\vec{e}_{31}   & \vec{e}_{32}   & \cdot    & \cdot    & \cdot    & \cdots & \cdot & \cdot \\
		\vec{e}_{41}   & \vec{e}_{42}   & \vec{e}_{43}   & \cdot    & \cdot    & \cdots & \cdot & \cdot \\
		\vec{e}_{51}   & \vec{e}_{52}   & \vec{e}_{53}   & \vec{e}_{54}   & \cdot    & \cdots & \cdot & \cdot \\
		\vdots   & \vdots   & \vdots   & \vdots   & \vdots   & \ddots & \vdots & \vdots \\
		\vec{e}_{H1}   & \vec{e}_{H2}   & \vec{e}_{H3}   & \vec{e}_{H4}   & \vec{e}_{H5}   & \cdots & \vec{e}_{H(H-1)} & \cdot
	\end{bmatrix}\;.
\end{align*}
Example edge attributes may be the physical distance between host pairs, time spent in close proximity, and their genetic distance.

The tensors $\mathbf{N}$ and $\mathbf{E}$ contain all of our epidemiological data, and in this structure can be thought of as a graph with $H$ nodes that is fully connected, with each node corresponding to a host. Node $i$ has attributes $\vec{n}_i$ embedding data specific to host $i$. Edge $(i,j)$ has attributes $\vec{e}_{ij}$, embedding relational data between hosts $i$ and $j$. Note that the edges have no directionality.

Our goal is to train a model on the dataset $\mathbf{N}$, $\mathbf{E}$ that can accurately predict whether any of the existing $H$ hosts in the dataset are closely related to some new host $H+1$. The genetic distance attribute appears in both $\mathbf{D}$, the object we are fitting, and $\mathbf{E}$, data used to fit the model. This is an important and unique aspect of the model structure: when making a prediction on the new host $H+1$ (the genetic distances between it and each of the $H$ existing hosts) that does not have a WGS sample, the model will make that prediction based on the attributes of host $H+1$, the attributes of the existing hosts $\mathbf{N}$, and also how those existing hosts relate to one another $\mathbf{E}$, including genetic distances.

\subsubsection*{Message passing}
\begin{figure}
	\centering
	\begin{tikzpicture}[	
		every node/.style={draw, minimum size=1cm, align=center, font=\sffamily},
		input/.style={circle, fill=inputcol},
		embed/.style={circle, fill=embedcol},
		output/.style={circle, fill=outputcol},
		square/.style={rectangle, minimum size=2cm, fill=modulecol},
		legend/.style={draw=none, fill=none, font=\sffamily\small},
		every path/.style={line width=1.5pt}
		]
		
		\node[input] (E) at (-0.2,0) {$\mathbf{E}$};
		\node[input] (N) at (-1.5,0) {$\mathbf{N}$};
		\node[input] (e) at (4.8,2.6) {$\vec{e}_{ij}$};
        
		\node[input, minimum size=1cm] (L1) at (-1,2) {$\vec{n}_{i}$};
		\node[square] (LSq) at (2,1.4) {\texttt{GeneralConv}};
		\node[embed] (LO) at (4,1.4) {$\tilde{\vec{n}}_{i}$};
		
		\node[input, minimum size=1cm] (R1) at (-1,-2) {$\vec{n}_{j}$};
		\node[square] (RSq) at (2,-1.4) {\texttt{GeneralConv}};
		\node[embed] (RO) at (4,-1.4) {$\tilde{\vec{n}}_{j}$};
		
		\node[embed, minimum size=1.5cm] (F) at (5.5,0) {$(\tilde{\vec{n}}_i, \tilde{\vec{n}}_j, \vec{e}_{ij})$};
		\node[square] (FinalSq) at (8.25,0) {\texttt{Linear}};
		\node[output] (Output) at (10.5,0) {$d^\text{pred}_{ij}$};
		
		\draw[->, bend left=15] (E) to (LSq);
		\draw[->, bend right=15] (E) to (RSq);
		\draw[->, bend left=25] (N) to (LSq);
		\draw[->, bend right=25] (N) to (RSq);
		
		\draw[->] (L1) -- (LSq);
		\draw[->] (R1) -- (RSq);
		
		\draw[->] (LSq) -- (LO);
		\draw[->] (RSq) -- (RO);
		
		\draw[->] (LO) -- (F);
		\draw[->] (RO) -- (F);
		\draw[->] (F) -- (FinalSq);
		\draw[->] (FinalSq) -- (Output);
		\draw[->] (e) -- (F);

		\begin{scope}[xshift=0cm, yshift=-4.5cm]
			\node[legend] at (0,2.0) {\textbf{}};
			\node[input, minimum size = 0.8cm, label=right:{\small Input data}] at (0,1.0) {};
			\node[square, minimum size = 0.8cm, label=right:{\small Neural network module}] at (5,1) {};
			\node[embed, minimum size = 0.8cm, label=right:{\small Embedded representation}] at (0,0) {};
			\node[output, minimum size = 0.8cm, label=right:{\small Final output}] at (5,0) {};
		\end{scope}
	\end{tikzpicture}
	\caption{Graph neural network architecture. This model evaluates the probability that a pair of hosts $i$, $j$ are closely related, when one of the hosts does not have a known pathogen sequence. The input data (light pink) are the node attributes of those individual hosts $\vec{n}_i$, $\vec{n}_j$, the edge attributes of that host pair $\vec{e}_{ij}$, the node attributes of all hosts in the dataset $\mathbf{N}$, and the edge attributes of all hosts in the dataset $\mathbf{E}$ (including known genetic distances). These data feed in through neural network modules (grey), with the intermediate outputs termed embedded representations (dark pink). The final output $d^{\mathrm{pred}}_{ij}$ (blue) is a scalar between 0 and 1.}
    \label{fig:tikz}
\end{figure}
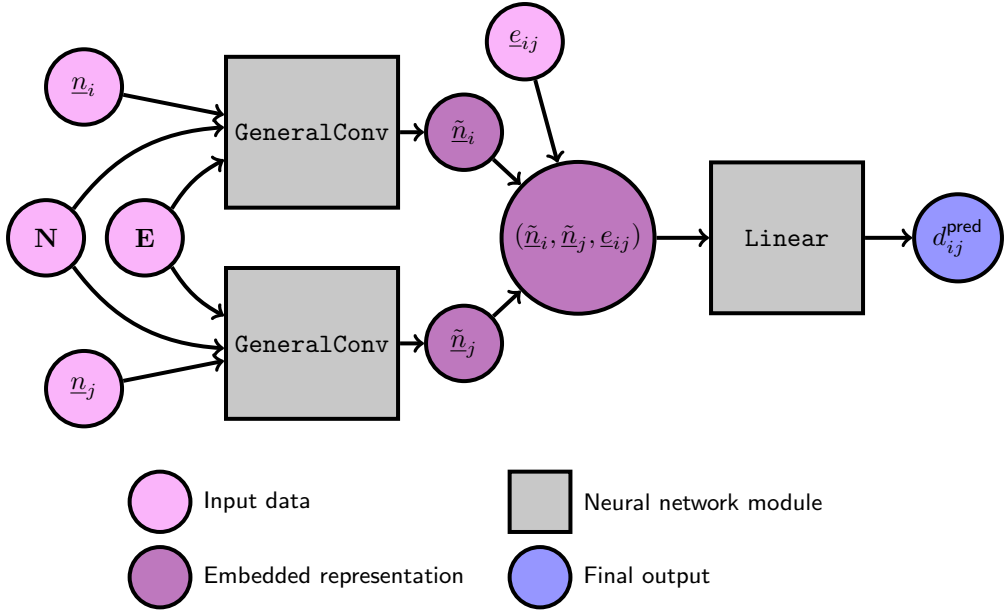

In this work we use the established Python neural network module $\texttt{conv.GeneralConv}$ within the package $\texttt{torch\_geometric}$ (v. 2.7.0). We describe the graph neural network architecture as originally outlined in Reference~\cite{you2020design} and the corresponding module~\cite{torch_geometric_generalconv} placed in the context of embedding epidemiological data, where nodes represent infected hosts. Figure~\ref{fig:tikz} shows for our application the model architecture and data flow.

For a given infected host $i$, their attributes are stored in the $(P\times 1)$-dimension vector $\vec{n}_i$. Rather than estimating genetic distances to other hosts from these attributes in isolation, we will first use the $\texttt{conv.GeneralConv}$ module to generate what is termed an embedding of node $i$, denoted $\tilde{\vec{n}}_i$. This is a vector representation of $i$ that will take into account the attributes of $i$, but also all other hosts $j \neq i$ relate to it. We want this embedding to be a vector of shape $(\lambda\times 1)$ where $\lambda$ is referred to as the number of output attributes.

First, we linearly transform $\vec{n}_i$ and each of its neighbours' attributes $\vec{n}_{j}$:
\begin{align*}
	\vec{n}_i &\mapsto \mathbf{W} \vec{n}_i + \vec{\text{B}} \\
	\vec{n}_j &\mapsto \mathbf{W'} \vec{n}_j + \vec{\text{B}}' \;.
\end{align*}
where $\mathbf{W}$ and $\mathbf{W}'$ are $(\lambda \times P)$ matrices, and $\vec{\text{B}}$ and $\vec{\text{B}}'$ (termed biases) are $(\lambda\times 1)$ vectors. These are learned parameters, whose values are fit in model training.

We then take into account the relationships between $i$ and all other nodes $j$, $\vec{e}_{ij}$. $\vec{e}_{ij}$ is a $(Q\times 1)$ vector. We transform the properties each edge $(i,j)$ as
\begin{align*}
	\vec{e}_{ij} \mapsto \mathbf{W''} \vec{e}_{ij} + \vec{\text{B}}''
\end{align*}
$\mathbf{W''}$ is a $(\lambda \times Q)$ matrix, $\vec{B}''$ is a $(\lambda \times 1)$ vector. These are also learned parameters.

For each neighbour $j$ of $i$, we then combine these transformations into a $(\lambda \times 1)$ vector $\vec{m}_{ij}$:
\begin{align*}
	\vec{m}_{ij} = \left(\mathbf{W} \vec{n}_i + \vec{\text{\text{B}}}\right) + \left(\mathbf{W'} \vec{n}_j + \vec{\text{B}}'\right) + \left(\mathbf{W''} \vec{e}_{ij} + \vec{\text{B}}''\right)\;,
\end{align*}
referred to as the ``message'' from $j$ to $i$. To ensure the embedding also includes the properties of $i$ itself, as a pragmatic solution we also include a message from a self-loop $m_{ii}$ where the edge-properties $\vec{e}_{ii} = \vec{0}$.

\subsubsection*{Attention}
The final embedding $\tilde{\vec{n}}_i$ of host $i$ is an aggregate of these messages from each of its neighbours. Each message is assigned a different weight, termed the message ``attention''. The purpose of attention is to allow the model to identify neighbouring nodes that are more or less important.

The attention given to each message is a scalar and is computed in an element-wise manner:
\begin{align*}
	\alpha_{ij}' &= \text{ReLU}\left(\vec{\text{A}}\cdot\vec{m}_{ij} \right) \\
	\alpha_{ij} &= \frac{\exp{(\alpha'_{ij})}}{\sum_{k\in\text{neighbours}(i)} \exp{(\alpha'_{ik})}}\;.
\end{align*}
where
\begin{align*}
	\text{ReLU}(x) = \text{Rectified Linear Unit}(x) = \text{max}(x,0)
\end{align*}
is an activation that introduces non-linearity into otherwise linear matrix operations. $\vec{\text{A}}$ is also a $(\lambda \times 1)$ vector, that is learned. In this instance the normalisation $\alpha'_{ij} \to \alpha_{ij}$ is a $\texttt{softmax}$ activation.  As a practical example of attention, if host $i$ is a bovid sampled in 2012, a bovid $j'$ sampled at the same holding at the same time is likely to provide more contextual information than a bovid $j''$ sampled in 2018 from an unconnected farm, so it is likely that the attention $\alpha_{ij'}>\alpha_{ij''}$.

\subsubsection*{Embedding}
The final embedding of node $i$ is then an aggregation of messages from each of its neighbours, weighted by their attention:
\begin{align*}
	\tilde{\vec{n}}_i = \sum_{j \in \text{neighbours}(i)}\alpha_{ij}\left[\mathbf{W} \vec{n}_i + \vec{B} + \mathbf{W'} \vec{n}_j + \vec{B}' + \mathbf{W''} \vec{e}_{ij} + \vec{B}''\right]
\end{align*}
resulting in a new, $(\lambda\times 1)$ vector informed by the attributes of node $i$ but also attributes of other nodes in the dataset. Unlike the original data in $\mathbf{N}$, $\mathbf{E}$, the values in the node embeddings are numerical outputs from a sequence of operations in the neural network module, and are not interpretable.

Node embedding can then be generalised in two ways. First, multiple message-passing layers can be stacked in sequence, where the input to layer $(k+1)$ is the output from layer $k$. This facilitates message passing beyond nearest neighbours and makes the neural network architecture deeper (a longer sequence of operations). Second, multiple parallel message passing processes can be computed with each one fit independently (termed multi-head attention). This makes the architecture wider (more operations in parallel) and is analogous to fitting different message passing opinions. In $\texttt{conv.GeneralConv}$ the embeddings from multiple attention heads are averaged.

\subsubsection*{From node embeddings to an estimated genetic distance}
From the message passing module we now have embedded representations $(\tilde{\vec{n}}_{i}, \tilde{\vec{n}}_{j})$, which we concatenate into a single vector. At this stage we also add the edge attributes of that host pair $\vec{e}_{ij}$, omitting the element that would correspond to the $\texttt{Genetic\_Distance}$ value if known. We introduce the edge attributes explicitly at this stage to ensure it more directly influences decision making, with the possibility of it being diminished if only included in $\mathbf{E}$. The concatenated embedding $\vec{l}_{ij} = (\tilde{\vec{n}}_{i}, \tilde{\vec{n}}_{j}, \vec{e}_{ij})$ (stacking the vectors) is a $((2\lambda+Q-1) \times 1)$ vector, and can be thought of as a representation of the explicit relations between $i$ and $j$ ($\vec{e}_{ij}$), and additional context on $i$ and $j$ when considering all other hosts in the dataset ($\tilde{\vec{n}}_{i}$, $\tilde{\vec{n}}_{j}$).

The data are then passed through a multilayer perceptron (MLP) module to output a single scalar value that represents the estimated genetic distance between host pairs. This is a standard neural network architecture, where input features undergo a set of linear transformations and activations~\cite{Goodfellow-et-al-2016}. An example of a 2-layer MLP could be a linear transformation by a $(8 \times (2\lambda+Q-1))$ matrix $\mathbf{M_1}$, followed by a transformation by a $(1\times 8)$ matrix $\mathbf{M_2}$:
\begin{align*}
	d_{ij}^\text{pred} = \text{sigmoid}\left(\mathbf{M}_2\,\left(\text{ReLU}\left(\mathbf{M}_1\,\vec{l}_{ij}\right)\right)\right)
\end{align*}
where the resulting value $d_{ij}^\text{pred}$ is a scalar that falls between 0 and 1. This is the probability that host $i$ and $j$ are closely genetically related. We take 1 as being the positive class (hosts $i$ and $j$ are closely genetically related) and 0 as the negative class (they are genetically distant).

\subsubsection*{Model training}
We train the model using backpropagation and gradient descent from the $\texttt{torch.optim}$ package~\cite{pytorch_optim}. This is an established package for training neural network models where model parameters are calibrated over successive iterations (epochs), to minimise a target loss function on the variable being fit. For the loss function we use binary cross-entropy loss on the final prediction probabilities $d_{ij}^\text{pred}$, adjusting for class imbalance. While fitting to the training set we monitor loss on the test set, and stop model training once test loss has plateaued.

\subsection{Pairwise models}
Alongside the GNN, we fit to the same underlying dataset a series of pairwise classification models: a logistic regression (LR) (\texttt{sklearn}, v. 1.6.0), random forest classifier (RF) (\texttt{sklearn}, v. 1.6.0)~\cite{breiman2001random}, and a boosted regression tree (BRT) (\texttt{xgboost}, v. 3.0.2)~\cite{chen2016xgboost}. These are all standard modelling methods for fitting an outcome variable (in this case the genetic distance $d_{ij}$ between hosts) based on the relational explanatory variables $\vec{e}_{ij}$ with the genetic distance attribute removed. The models are trained on pairwise data and do not explicitly learn higher-order relations. RF and BRT are tree-based models which with appropriate hyperparameter selection are largely robust to correlations in the explanatory variables, and are capable of learning complex, nonlinear multivariate relationships between the explanatory variables and the outcome variable.

\subsection{Model training and hyperparameter selection}
Explanatory variables are normalised across all models such that each explanatory variable has mean zero, and standard deviation 1.

For the GNN we include edge-level and node-level attributes that may be derived from the same raw data (e.g.~including $\texttt{xcoord}$ and $\texttt{ycoord}$ as node properties for each host, as well as $\texttt{physical\_distance}$ as an edge property). This pragmatic approach introduces a level of redundancy but allows the GNN to freely learn from both representations of the data.

We use $70\%$ of nodes for model training. The train/test allocation is random and test nodes are not used in the hyperparameter selection process. The same train/test allocation is used for each model type. To identify the best hyperparameters for each modelling architecture, we fit an ensemble of models using 5-fold cross-validation across different hyperparameter combinations, seeking to minimise binary cross-entropy loss on the unseen validation portion of the data. For host pair classification, unseen data would be a new host $H+1$ that has life history data but no pathogen sequence. In the model, this would be a node $H+1$ with $H$ new edges between it and the existing $H$ hosts. We then train each model on $5$ folds where each fold has $0.2H'$ nodes removed, and assess model performance based on the genetic distance between the $0.8H'$ trained nodes and the $0.2H'$ unseen nodes. We choose from the ensemble of models the hyperparameter combination with the highest balanced accuracy on the unseen validation nodes.

For each model type we then fit a model using the best hyperparameters. Model performance is the evaluated on predictive performance on test edges (edges between train hosts and unseen test hosts, giving $0.7H\times0.3H$ test edges in total).

\subsection{Model performance}
As a classification task with large class imbalance, an effective model is one that can correctly identify closely-related host pairs (true positives, TP) while minimising the host pairs incorrectly identified as close when they are in reality distant (false positives, FP). We quantify performance by balanced accuracy (BA, arithmetic mean of positive-class recall and negative-class recall, equally weighting both classes) on test edges:
\begin{equation*}
	\text{BA} = \frac{1}{2}\left(\frac{\text{TP}}{\text{TP}+\text{FN}}+\frac{\text{TN}}{\text{TN}+\text{FP}}\right)
\end{equation*}
with TN and FN true and false negatives. The detection threshold is fixed at 0.5 (if $d_{ij}^\text{pred} > 0.5$, the host pair is classified as close). As an additional threshold-free measure also record the ROC-AUC (area under the receiver-operator characteristic curve). A purely random model would yield a BA and ROC-AUC of $0.5$. 

To quantify model confidence we calculate the mean prediction entropy (MPE)
\begin{equation*}
	\text{MPE} = -\frac{1}{\frac{1}{2}H(H-1)}\sum_{\text{all pairs $i,j$}} d^\text{pred}_{ij}\log{d^\text{pred}_{ij}} + (1-d^\text{pred}_{ij})\log{(1-d^\text{pred}_{ij})}\;.
\end{equation*}
This is a measure of the distribution of model predictions over all test edges. Models producing more confident predictions will have $d^\text{pred}_{ij}$ values clustering around 0 and 1 and a lower MPE. Conversely, less confident models will return a broader distribution (more $d^\text{pred}_{ij}$ around 0.5), and a higher MPE. We stress that MPE reflects confidence in model predictions, rather than accuracy of those predictions.

\subsection{Explanatory variable importance}
To quantify the explanatory power of different variables across different models types in a consistent manner we calculate a permutation-based importance. For each explanatory variable (e.g.~physical distance, time spent in close proximity), we randomly permute its values in the dataset (effectively removing it) and generate a new set of model predictions in this modified dataset. The importance of a given variable is defined as the change in balanced accuracy on test edges using this permuted dataset. For the GNN we calculate importance of both node-level attributes and edge-level attributes. In particular for the GNN the importance of the $\texttt{Genetic\_Distance}$ variable, the known genetic distances between existing hosts, indicates whether information on existing host pairs improves the model's predictive power on classifying new host pairs.

\section{Results}\label{sec:Results}
\begin{table}[t]
    \centering
    {\sffamily
    \footnotesize
    \begin{tabularx}{11cm}{|p{2cm}|p{1.5cm}|Y|Y|Y|}
        \hline
		\textbf{Dataset} & \textbf{Model} & \textbf{Balanced accuracy (BA)} & \textbf{ROC-AUC} & \textbf{Mean precision entropy (MPE)} \\
        \hline
		\multirow{4}{*}{\textbf{Synthetic 1}}
            & \textbf{BRT} & 0.680 & 0.745 & 0.561 \\
            & \textbf{RF}  & 0.602 & 0.738 & 0.352 \\
            & \textbf{LR}  & 0.657 & 0.715 & 0.632 \\
            \cdashline{2-5} 
            & \textbf{GNN} & 0.798 & 0.871 & 0.444 \\
        \hline
		\multirow{4}{*}{\textbf{Synthetic 2}} 
            & \textbf{BRT} & 0.670 & 0.740 & 0.493 \\
            & \textbf{RF}  & 0.574 & 0.729 & 0.290 \\
            & \textbf{LR}  & 0.666 & 0.728 & 0.601 \\
            \cdashline{2-5} 
            & \textbf{GNN} & 0.807 & 0.869 & 0.452\\
        \hline
		\multirow{4}{*}{\textbf{Synthetic 3}}
            & \textbf{BRT} & 0.632 & 0.684 & 0.608 \\
            & \textbf{RF}  & 0.616 & 0.681 & 0.515 \\
            & \textbf{LR}  & 0.623 & 0.669 & 0.644 \\
            \cdashline{2-5} 
            & \textbf{GNN} & 0.743 & 0.853 & 0.381 \\
		\hline
		\multirow{4}{*}{\textbf{Woodchester}}
            & \textbf{BRT} & 0.624 & 0.858 & 0.079 \\
            & \textbf{RF}  & 0.546 & 0.872 & 0.141 \\
            & \textbf{LR}  & 0.798 & 0.866 & 0.419 \\
            \cdashline{2-5} 
            & \textbf{GNN} & 0.789 & 0.860 & 0.484 \\
		\hline
		\multirow{4}{*}{\textbf{Cumbria}}
            & \textbf{BRT} & 0.647 & 0.733 & 0.238 \\
            & \textbf{RF}  & 0.641 & 0.742 & 0.199 \\
            & \textbf{LR}  & 0.617 & 0.686 & 0.623 \\
            \cdashline{2-5} 
            & \textbf{GNN} & 0.709 & 0.764 & 0.556 \\
		\hline
	\end{tabularx}
    }
	\caption{Performance metrics by dataset and model type, comparing mean prediction entropy (where lower values indicate more confident model predictions, but not necessarily more accurate), area under the receiver-operator characteristic curve (ROC-AUC, threshold-free measure of accuracy) and balanced accuracy (accuracy accounting for class imbalance).}
	\label{tab:performance_metrics_models}
\end{table}
Table~\ref{tab:performance_metrics_models} summarises model performance across each of the datasets. The final GNN architectures for each dataset are given in Supplementary Material, Section~\ref{sec:GNNarchitectures}.

\begin{figure}
	\includegraphics[width=\textwidth]{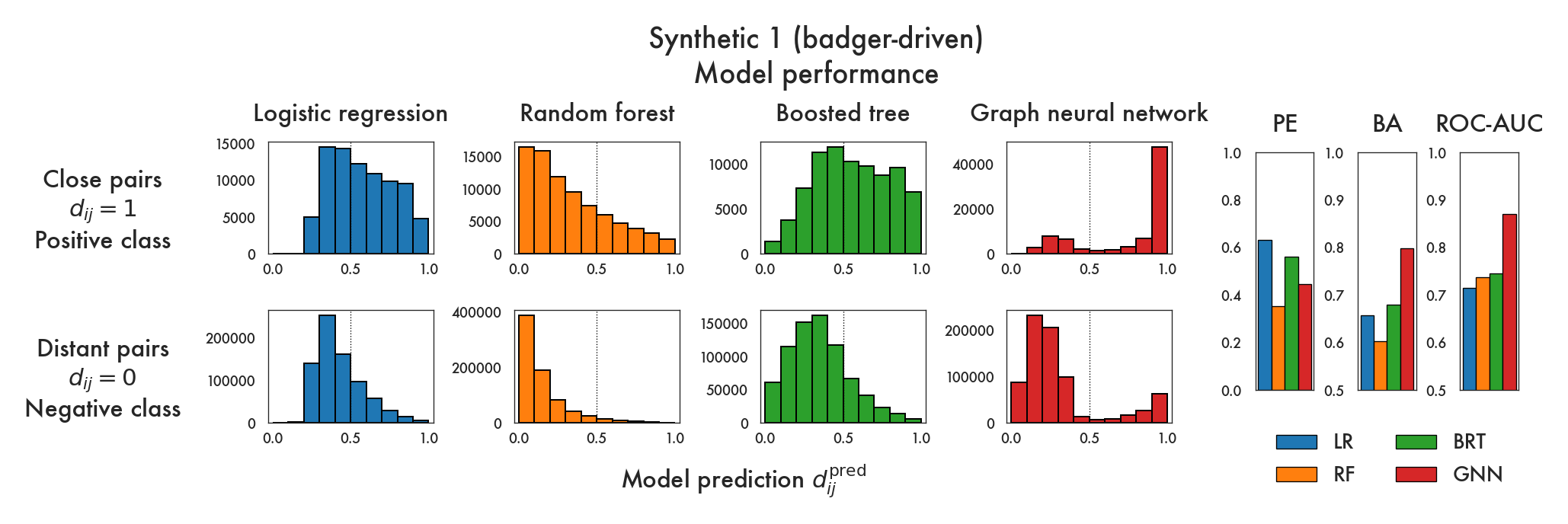}
	\includegraphics[width=\textwidth]{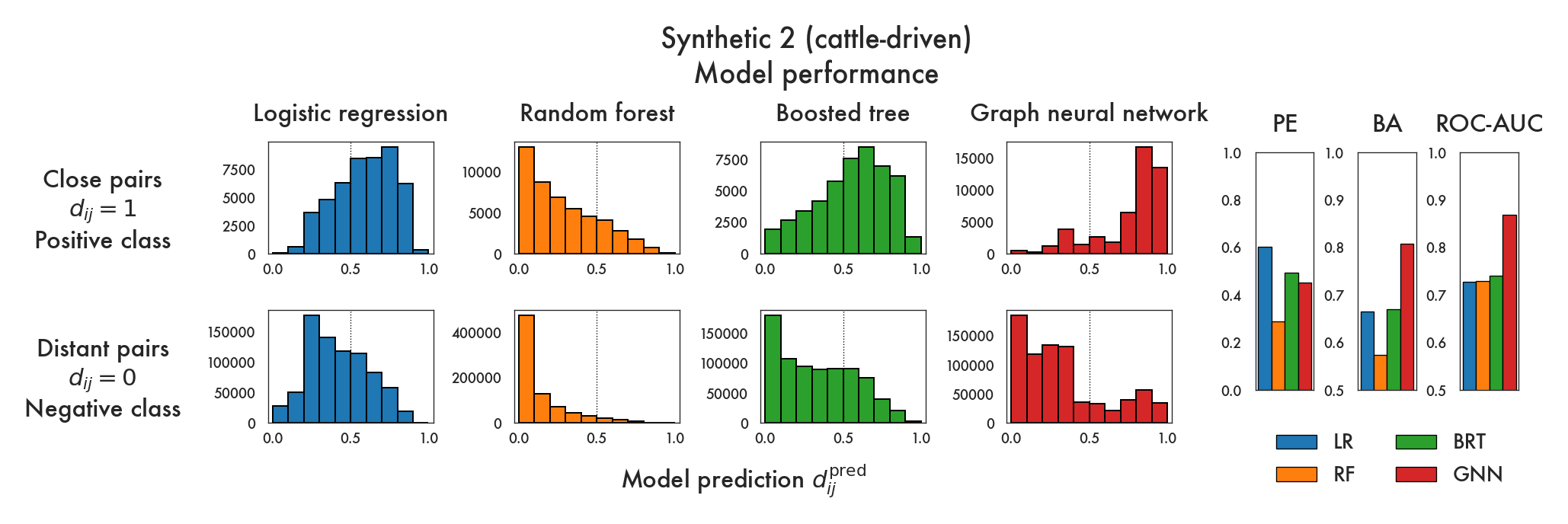}
	\includegraphics[width=\textwidth]{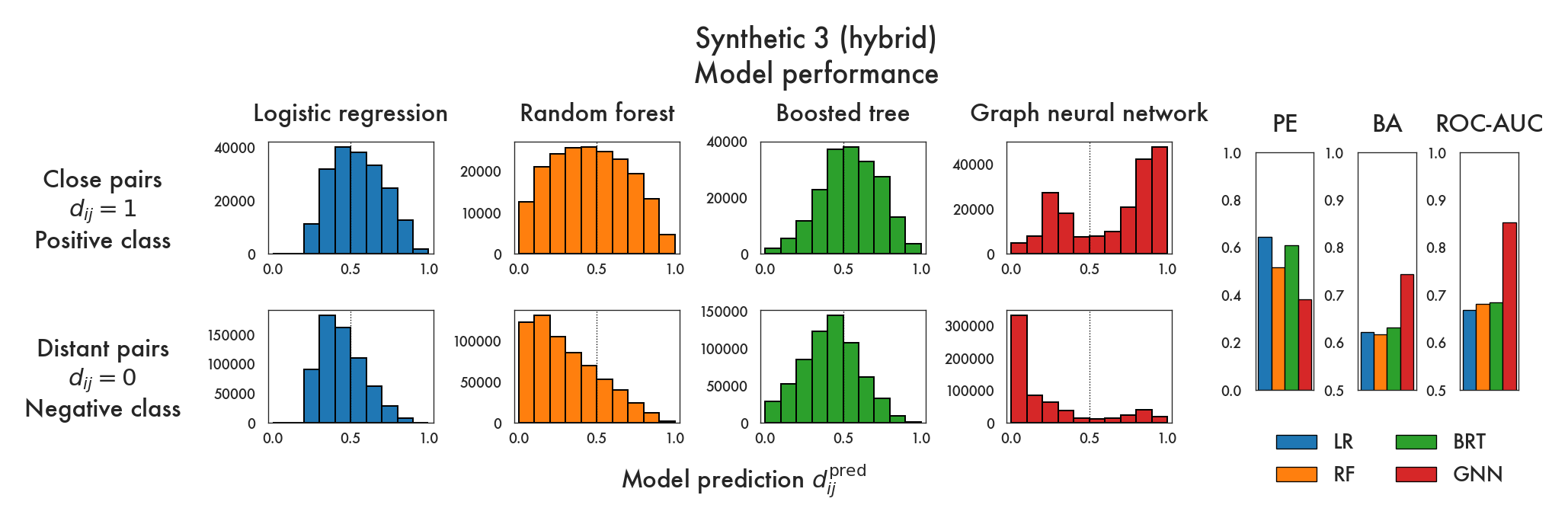}
	\caption{Model performance over the synthetic datasets ($H=2\,000$ hosts each). Left: Classification of test host pairs $(i,j)$ for each model, separated by whether they are truly closely related $(d_{ij} = 1)$ or distant $(d_{ij} = 0)$. Right: mean prediction entropy (MPE, where lower MPE indicates more confident predictions), balanced accuracy (BA) and area under the receiver-operator characteristic curve (ROC-AUC).}
	\label{fig:performance_S}
\end{figure}

\begin{figure}
	\includegraphics[width=\textwidth]{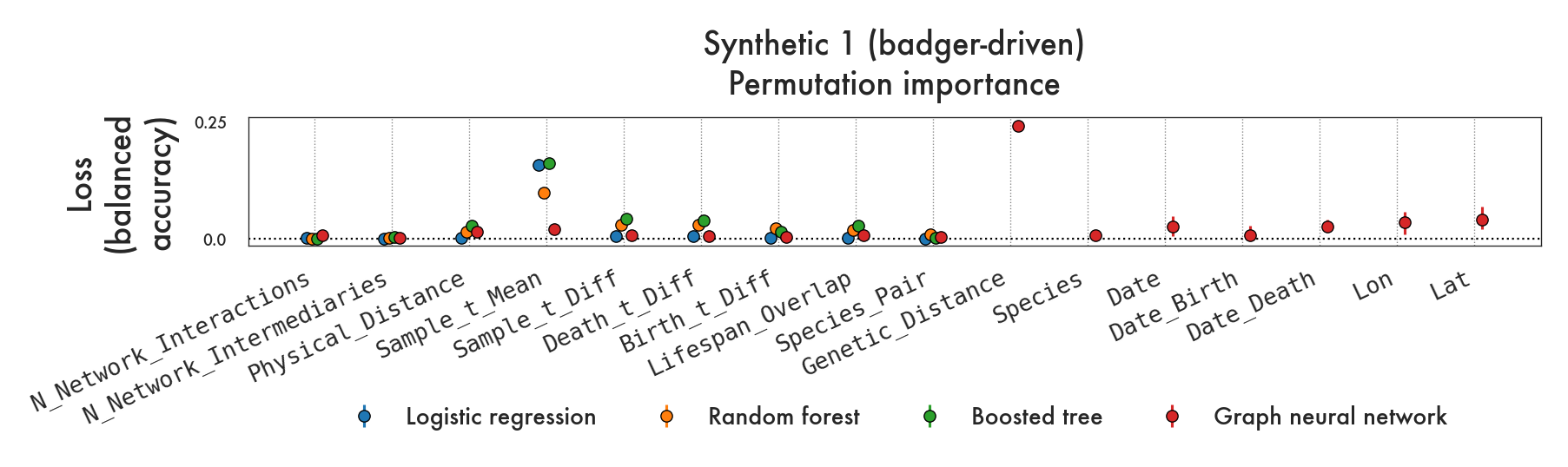}
	\includegraphics[width=\textwidth]{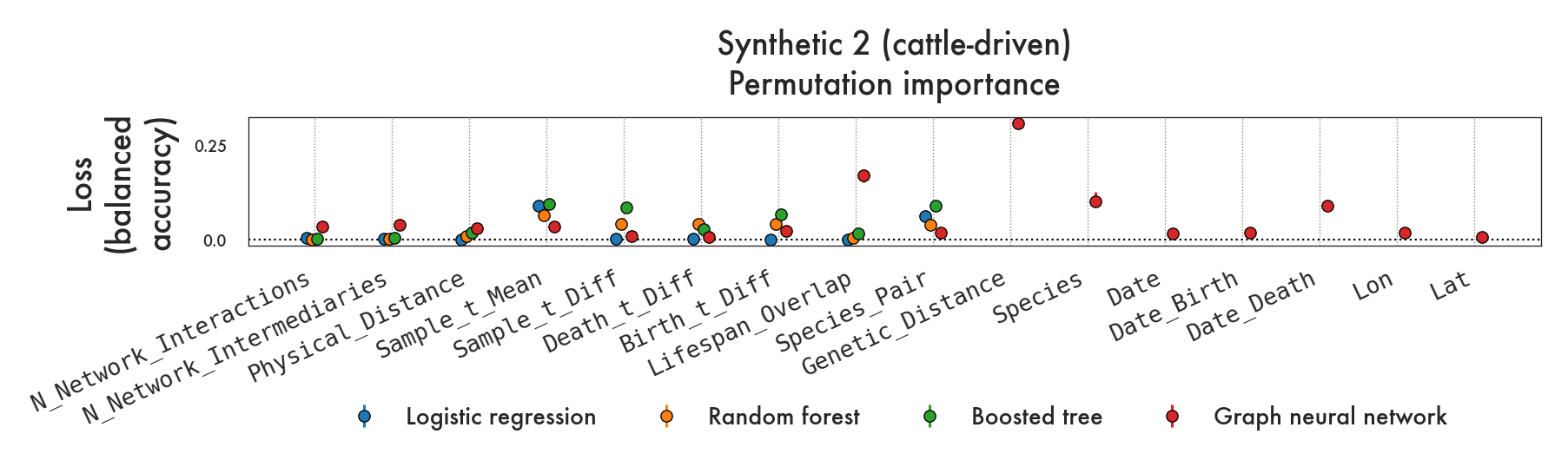}
	\includegraphics[width=\textwidth]{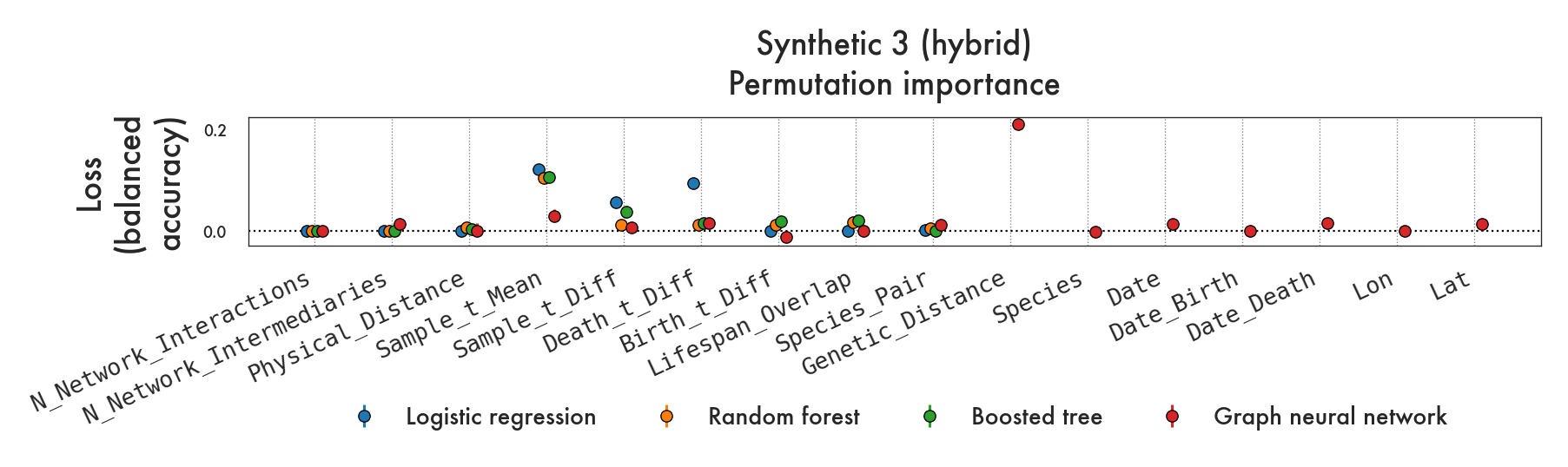}
	\caption{Model importance over the synthetic models. This is the loss in balanced accuracy on the test host pairs, on random permutation of a given variable in the dataset (effectively removing it). The $\texttt{Genetic\_Distance}$ attribute is only populated for edges in the train dataset for the GNN model. Other variables with a value for the GNN only are node-level attributes.}
	\label{fig:importance_S}
\end{figure}

We first evaluate performance on the three synthetic datasets before considering the smaller real-world datasets. As quantified by BA (0.798, 0.807, 0.743 respectively) and ROC-AUC (0.871, 0.869, 0.853), the GNN models have high intrinsic discriminatory power, especially as compared to the three pairwise models. Figure~\ref{fig:performance_S} shows the GNNs achieve convincing class separation, and can be characterised as highly sensitive by correctly and confidently classifying a higher proportion of close pairs when one host was not previously seen by the model. The proportions of distant host pairs incorrectly classified as close are small, but high in number due to the significant imbalance of classes in these data. Evaluating explanatory variable importance, Figure~\ref{fig:importance_S} indicates that the $\texttt{Genetic\_Distance}$ attribute (recalling this is on train edges, available to the GNN only) is highly important relative to other variables thus substantially influences how the GNN makes predictions on unseen hosts.

\begin{figure}
	\includegraphics[width=\textwidth]{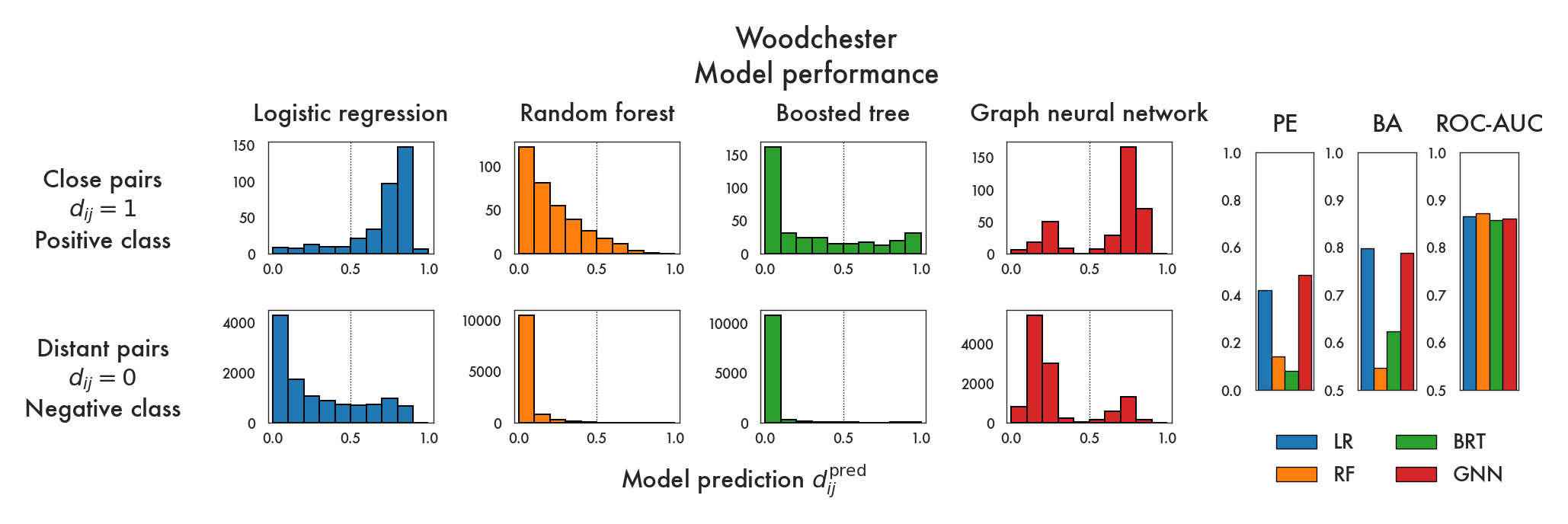}
	\caption{Model performance over the Woodchester dataset ($H=241$ hosts). Left: Classification of test host pairs $(i,j)$ for each model, separated by whether they are truly closely related $(d_{ij} = 1)$ or distant $(d_{ij} = 0)$. Right: mean prediction entropy (MPE, where lower MPE indicates more confident predictions), balanced accuracy (BA) and area under the receiver-operator characteristic curve (ROC-AUC).}
	\label{fig:performance_W}
\end{figure}

\begin{figure}
	\includegraphics[width=\textwidth]{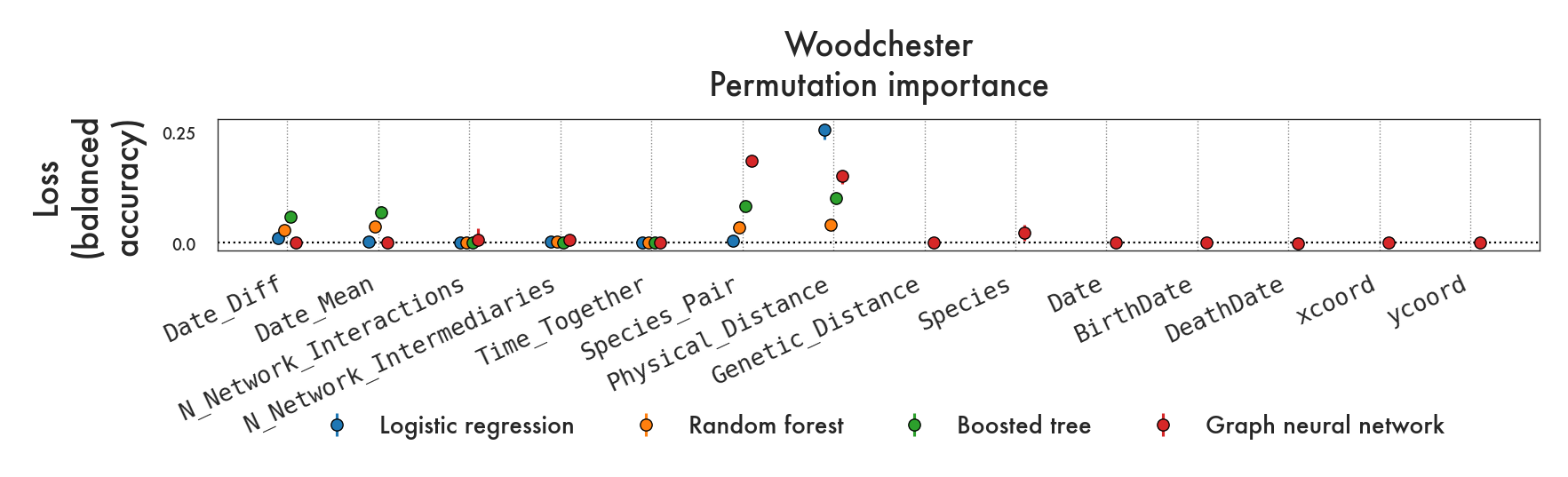}
	\caption{Model importance over the Woodchester model as quantified by the loss in balanced accuracy on the test host pairs, on random permutation of a given variable in the dataset (effectively removing it). The $\texttt{Genetic\_Distance}$ attribute is only populated for edges in the train dataset for the GNN model. Other variables with a value for the GNN only are node-level attributes.}
	\label{fig:importance_W}
\end{figure}

Performance on the smaller Woodchester dataset (Figure~\ref{fig:performance_W}, where host pairs $\leq4$ SNPs difference are labelled as close) is mixed. First, the much simpler logistic regression demonstrates comparable accuracy to the GNN on both BA (0.798, 0.789 respectively) and ROC-AUC (0.866, 0.860). Despite cross-validation and training under a loss function correcting for class imbalance, the random forest and boosted tree models are poor at distinguishing close host pairs above the detection threshold of $0.5$. While a lower detection threshold may improve accuracy (as indicated by the threshold-free ROC-AUC) the optimal choice is unknowable under real-world conditions. Feature importance in Figure~\ref{fig:importance_W} shows that across all models the most important explanatory variables are $\texttt{Species\_Pair}$ $\texttt{Physical\_Distance}$, $\texttt{Date\_Diff}$ and $\texttt{Date\_Mean}$, and for the GNN the $\texttt{Genetic\_Distance}$ attribute offers no statistically significant explanatory power.

\begin{figure}
	\includegraphics[width=\textwidth]{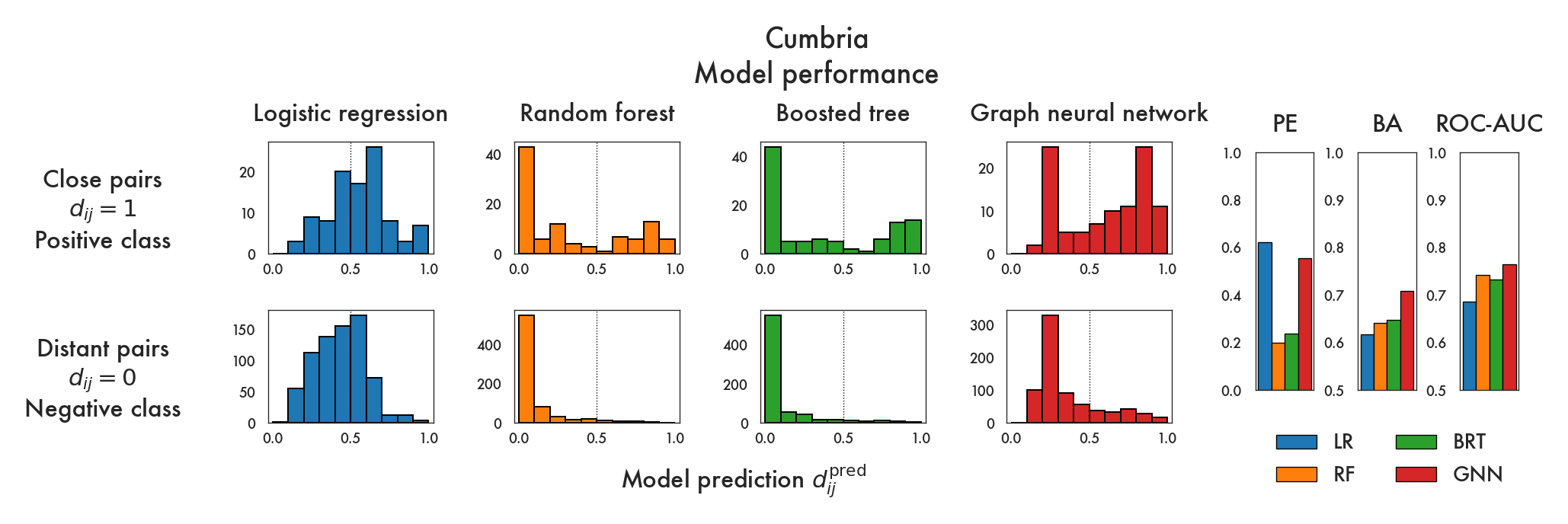}	
	\caption{Model performance over the Cumbria dataset ($H=63$ hosts). Left: Classification of test host pairs $(i,j)$ for each model, separated by whether they are truly closely related $(d_{ij} = 1)$ or distant $(d_{ij} = 0)$. Right: mean prediction entropy (MPE, where more confident predictions have lower entropy), balanced accuracy (BA) and area under the receiver-operator characteristic curve (ROC-AUC).}
	\label{fig:performance_C}
\end{figure}

\begin{figure}
	\includegraphics[width=\textwidth]{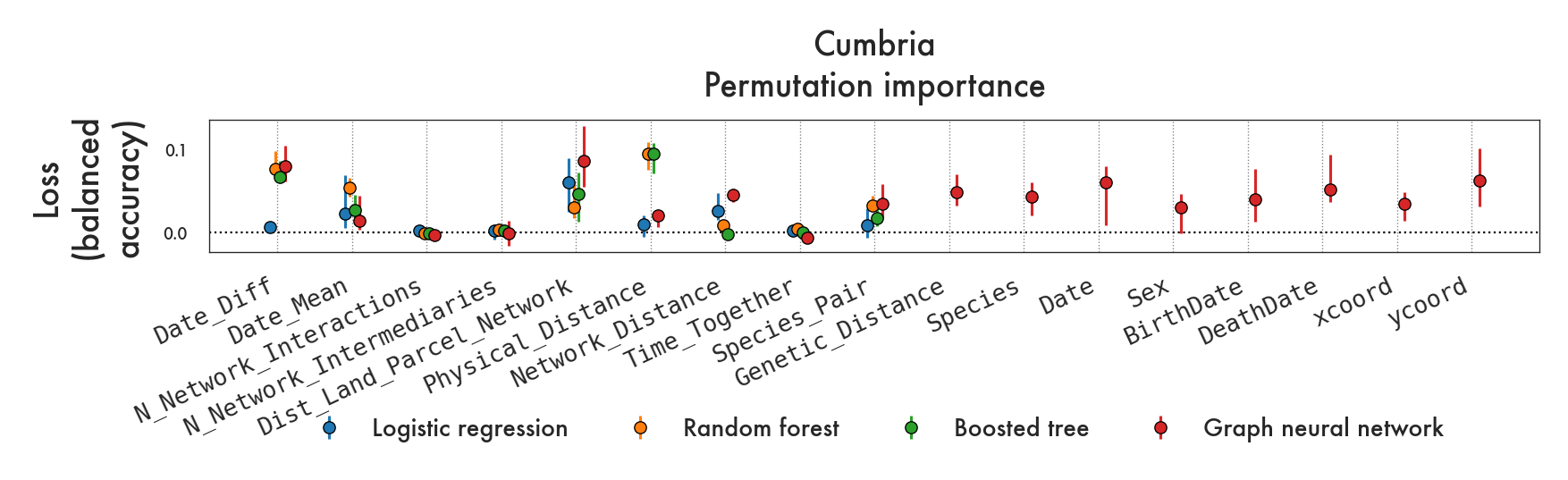}	
	\caption{Model importance over the Cumbria model as quantified by the loss in balanced accuracy on the test host pairs, on random permutation of a given variable in the dataset (effectively removing it). The $\texttt{Genetic\_Distance}$ attribute is only populated for edges in the train dataset for the GNN model. Other variables with a value for the GNN only are node-level attributes.}
	\label{fig:importance_C}
\end{figure}

On the even smaller Cumbria dataset model predictive power is relatively poor across all model types ($\text{ROC-AUC} = 0.686-0.764$, $\text{BA} = 0.617-0.709$; for both metrics random classification scores 0.5 and perfect classification scores 1.0). Figure~\ref{fig:performance_C} indicates that none of the models achieve convincing class separation, including the GNN. In contrast with the Woodchester dataset, the $\texttt{Genetic\_Distance}$ edge variable does have statistically significant additional explanatory power (Figure~\ref{fig:importance_C}).

\section{Discussion}\label{sec:Discussion}
\subsubsection*{Graph neural networks are a natural architecture for precision epidemiology problems}
Precision epidemiology aims to identify broad-scale risk factors for infectious disease transmission, and ultimately inform control strategies, by inferring specific transmission pathways and infector-infectee pairs. Because of the essential nature of infectious disease spread, the samples and associated metadata gathered for these tasks are not independent observations, but are intrinsically interconnected.

Our aim in this methodological study was to demonstrate how graph neural networks (GNNs) are a natural and flexible modelling architecture for making insights from such interconnected datasets. To predict the relationship between two hosts, GNNs use data on those hosts, but are also free to leverage the whole dataset for additional contextual detail (such as how those two hosts are related to other hosts). This contrasts with traditional pairwise regression approaches where data are represented as a set of pairwise observations that are treated as independent.

\subsubsection*{Model performance}
While the methodology is generally applicable to any infectious disease dataset with a relational attribute of interest, we have fit GNNs to bovine Tuberculosis (bTB) disease datasets in Great Britain containing information on sample time, locations and host life histories, \emph{M. bovis} pathogen whole-genome sequences. The task was to classify whether a given pair of hosts' pathogen sequences are closely genetically related, as a proxy for being closely related in the overall transmission tree.

When using synthetic datasets (each with $H=2\,000$ hosts sampled from a comprehensive multispecies bTB model simulating real-world cattle movements), GNNs achieve convincing class separation when predicting relations to new hosts (i.e.~the test set) as compared to pairwise models (logistic regression, boosted regression trees, random forests). Across these three datasets the $\texttt{Genetic\_Distance}$ attribute (the known genetic distances between existing hosts in the dataset) emerged as a highly important variable for model prediction, as quantified by balanced accuracy loss of model after randomly permuting that variable. This indicates that the additional context provided by the known genetic distances between other host pairs in the dataset) contributed significantly to the model's predictive power.

We then trained GNNs on two real-world datasets, Woodchester Park ($H = 241$) and Cumbria ($H = 63$). These datasets were much smaller than the synthetic datasets, particularly in terms of edges which scale quadratically with sample size. For these, classification power was unconvincing over all model types including the GNN. 

The Woodchester dataset contained 130 sequenced cattle as compared to approximately $~6\,000$ confirmed infections in the study period~\cite{wood2024utility}. It is difficult to determine an exact coverage, but as not all infected cattle are tested (and if tested may not necessarily be test-sensitive) it is reasonable to assume that cattle coverage is below 1\%. As wild animals, badger coverage is even more difficult to estimate, but in the study period where the 111 badger WGS samples were gathered, $~$700 badgers were found with confirmed infection. Of the $28\,920$ pairs, the median genetic distance was $26$ SNPs and only $91$ had a genetic distance of $0-1$ SNPs. With the substitution rate of \emph{M. bovis} being of order $0.1-1$ substitutions per genome per year, the practical insight to be gained from beyond-pairwise relations with samples of such high diversity may therefore be limited, consistent with the $\texttt{Genetic\_Distance}$ attribute emerging as unimportant in the GNN. In addition, as Woodchester Park is surrounded by a much larger region with sustained, endemic bTB transmission, there is potential for hosts in the dataset to have been infected via external (and unknown to the model) transmission pathways, entirely unrelated to other hosts in the data. With these factors it may then be challenging for any approach to produce a convincing predictive model from such data.

The Cumbria dataset has fewer hosts still, but differs from Woodchester Park in that it is from a contained outbreak over a shorter time interval. In this case, the edge-level $\texttt{Genetic\_Distance}$ attribute did improve model predictive power to a statistically significant degree suggesting that the GNN is able to make improved predictions when leveraging the full dataset as compared to the other models. However in the context of limited data (with a 30\% test split yielding a test set of 18 hosts) model performance metrics are inherently highly variable, and sensitive to train-test allocation of nodes as well as hyperparameter choice.

More generally, across all datasets and model types a substantial number of false positives were classified (genetically distant hosts that were incorrectly identified as close) relative to true positives. Given the nature of the modelling task these are difficult to avoid entirely; a pair of infected hosts may have life history data consistent with being closely genetically related, but simply happen to not be related (as explored by the authors in Reference~\cite{wood2024utility}, in the context of quantifying the additional explanatory power provided by sequence data). A practical example would a pair of infected badgers that tested positive for bTB at the same time and location, but happened to have been infected from different sources. In principle this subset of host pairs are be where the GNN may be able to leverage additional context (such as, for this example, the genetic diversity of surrounding badgers) to correctly identify these negatives.

\subsubsection*{Extensions, limitations, model architectures}
For simplicity of model fitting and evaluation we have treated genetic distance estimation in this work as a classification task; predicting if two hosts are closely related or not. We stress that a more detailed regression task (e.g.~rather than predicting a binary close/not-close relation, predicting the exact number of SNPs between two host's pathogens) is a simple extension that would require interpreting the final value $d_{i,j} \in [0,1]$ as, for example, $d_{ij} = 1/(1+\text{SNPs}_{ij} )$, and fitting using a different loss function suitable for regression. Finally, models that consider the relative branch lengths from the MRCA (i.e. a graph where each host pair is connected by two edges, one in each direction) would allow for greater discrimination---separating out, for example, transmission chains leading from one sample to another, from circumstances where the two samples are infected by a common source.

As with machine learning models in general, a key limitation to GNN approaches over traditional statistical models is in model interpretability; understanding why a host pair has been classified as close or distant. While less standard in the GNN code base, we have probed the importance of explanatory variables by the accuracy loss of the model when effectively removing that variable by random permutation. Directionality (how increasing or decreasing a variable's value affects the model prediction, analogous to a gradient in traditional models) could in principle also be probed in GNNs by methods analogous to partial dependencies~\cite{friedman2001greedy} or accumulated local effects~\cite{apley2018package}. Model uncertainty is also difficult to interpret here; while the output of all these models $d_{ij}^{\text{pred}}\in[0,1]$ these values should not necessarily be interpreted as a probability with embedded model uncertainty.

While beyond the scope of the analysis done here, an advantage of graph neural networks relevant to precision epidemiology is that hosts with incomplete metadata may nonetheless be included as additional nodes in the input graph used to train the model. In this regard GNNs may be particularly effective when seeking to integrate all available data from an infectious disease outbreak of varying quality and completeness, into a unified statistical model. An example relevant to bTB would be, for a farm bTB outbreak where only a single cow is sequenced, including all reactors from that breakdown, sequenced or not, as connected nodes. While those reactors would not have an associated $\texttt{Genetic\_Distance}$ attribute, their relations (locations, physical interactions) with other hosts in the dataset are nonetheless available to the model during training, and in making predictions.

Given the countless number of ways neural network modules can be layered, our modelling choices in this work were pragmatic in nature and by no means exhaustive. We have primarily used the $\texttt{conv.generalConv}$ module~\cite{torch_geometric_generalconv} as it explicitly uses edge attributes alongside node attributes in message passing (other graph modules such as $\texttt{GATConv}$~\cite{velivckovic2017graph}, $\texttt{GATv2Conv}$~\cite{brody2021attentive} and $\texttt{GCNConv}$~\cite{kipf2016semi} either do not use multiple edge attributes, or only use edge attributes for determining attention weights only), followed by a standard set of linear transforms to reduce dimensionality to a scalar value. More effective modules and architectures tailored to epidemiological datasets, specifically those that include edge attributes more explicitly in later layers, may well exist.

Finally, the modelling task here was at edge-level; fitting relations between nodes in a graph. We finish by noting that GNNs are highly flexible and could also be structured to address other epidemiological problems. Possible examples include: classifying the probability that one host infected another (an edge-level task with directed edges); classifying some overall property of an infectious disease dataset (a graph-level task); or, for each host, a probability that it was the index case (a node-level task). For these tasks the structure of the model and data flow may differ, but it would still pass messages between nodes representing infected hosts and allow use of the full dataset to make any model prediction.

To conclude, we have demonstrated how GNNs can leverage infectious disease whole-genome sequence data to their fullest to uncover underlying patterns of transmission, by treating the data as an intrinsically interconnected graph. GNNs exhibit has superior performance in predicting host-host relationships compared to other traditional methods we tested, albeit decreasingly so with smaller datasets where there is less information to be gained beyond first-order host-host relationships. Our implementation illustrates a flexible framework that could be highly effective for extracting insights from larger, more comprehensive epidemiological datasets.

\clearpage

\clearpage

\clearpage


\clearpage

\section{Code availability}
The code base is written in Python version 3.10.16 and is run on macOS (arm64) 15.6.

Underlying code is available at \url{https://git.ecdf.ed.ac.uk/awood310/genetic-distance-graph-neural-networks/}.

\bibliographystyle{vancouver}
\bibliography{gnn}

\newpage
\setcounter{page}{1}

\section*{\LARGE{Supplementary Material}} 

\emph{\large{Learning relationships in epidemiological data using graph neural networks}} \\

A. J. Wood, A. R. Sanchez, R. R. Kao

\clearpage

\appendix

\renewcommand\thefigure{S\arabic{figure}}    
\renewcommand\thetable{S\arabic{table}}    
\setcounter{figure}{0}  
\setcounter{table}{0}  

\section{Data details}\label{supp:Data}

\begin{figure}[h]
	\includegraphics[width=\textwidth]{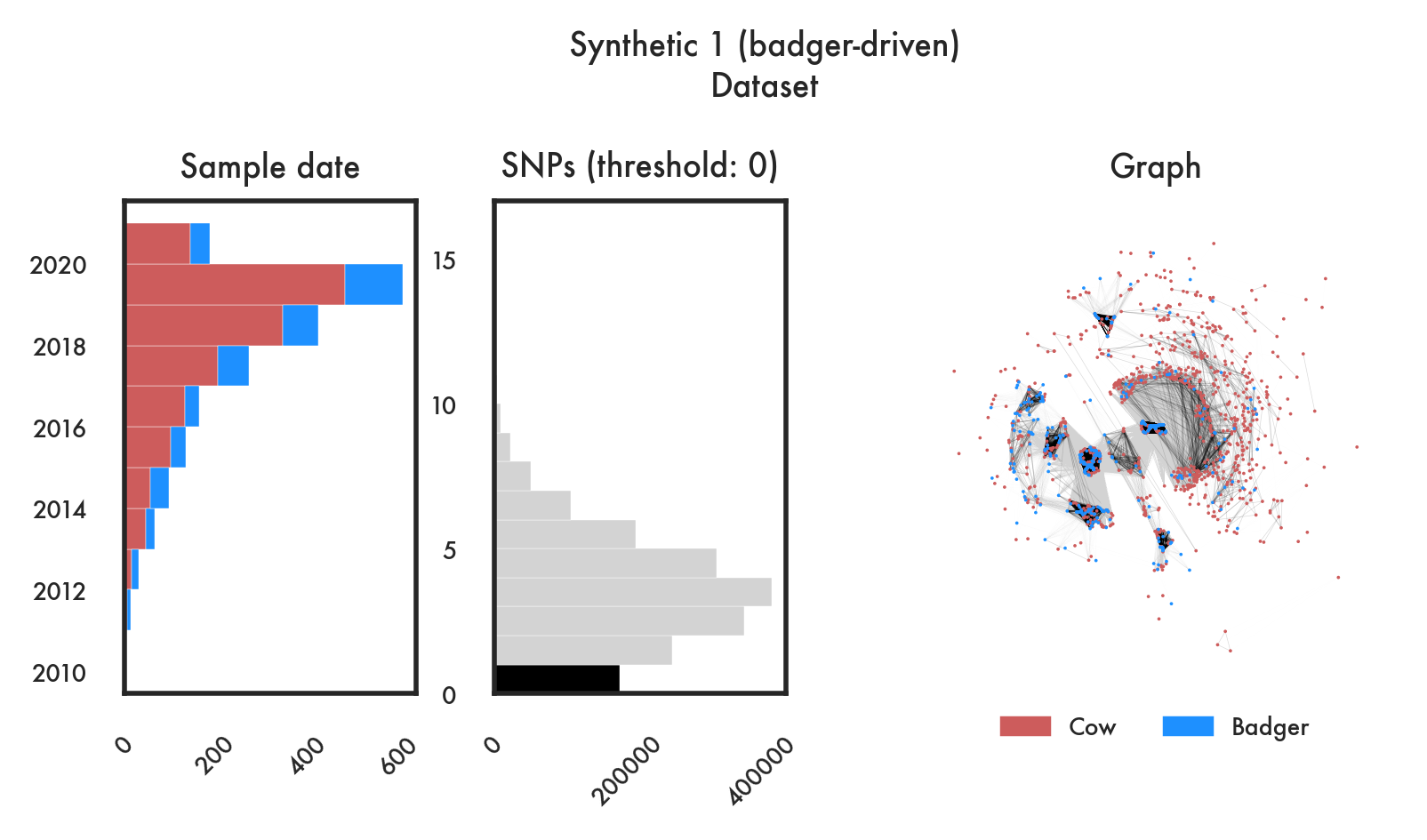}
	\caption{Summary of the Synthetic 1 dataset, run under simulation dynamics where cattle infections primarily come from badgers. Left: time distribution of samples broken down by species. Centre: distribution of pairwise genetic distances. Right: graph representation where genetically similar hosts are clustered. Host pairs classified as being close are highlighted in black.}
	\label{fig:dataS1}
\end{figure}

\begin{figure}
	\includegraphics[width=\textwidth]{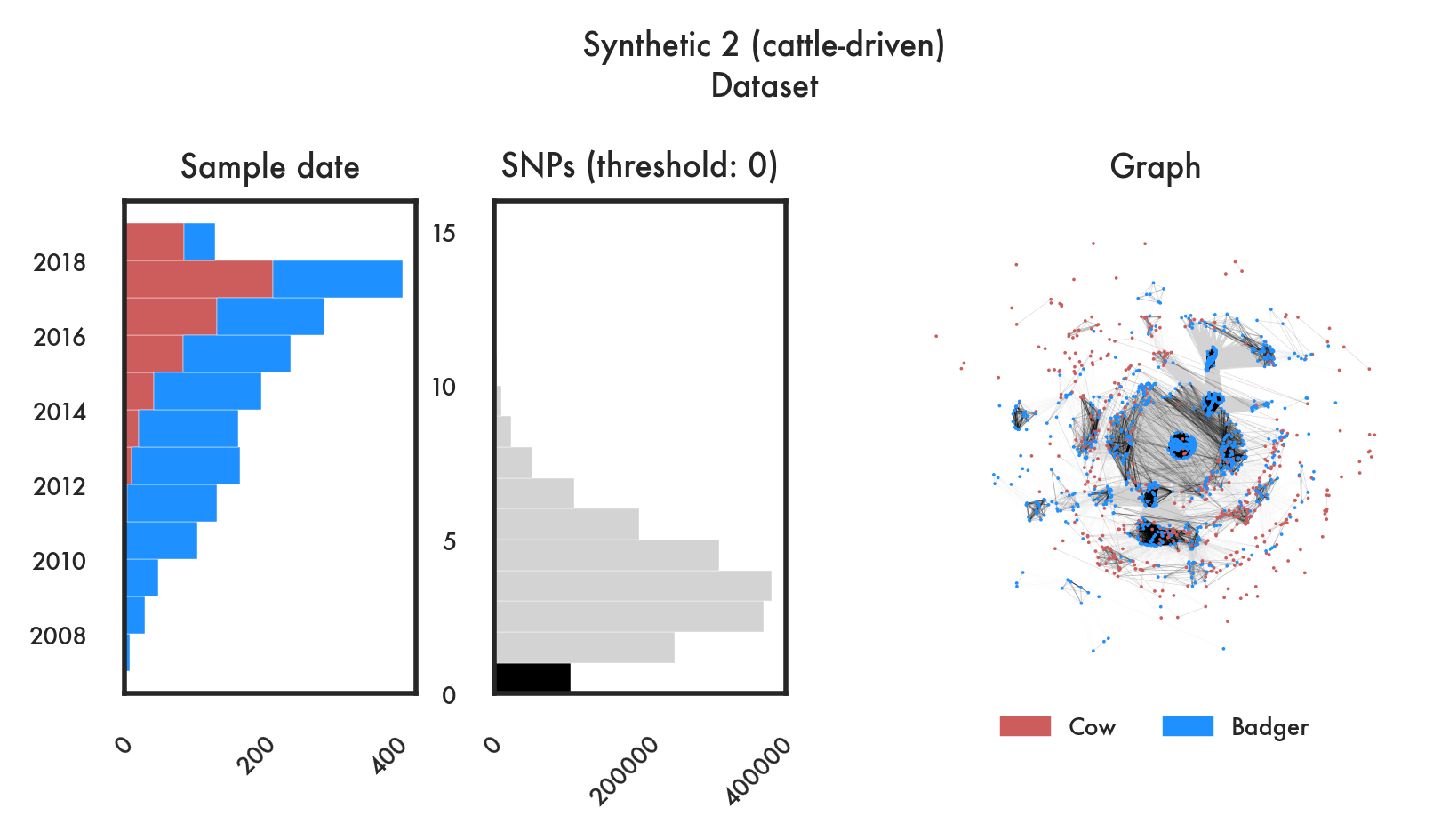}
	\caption{Summary of the Synthetic 2 dataset, run under simulation dynamics where cattle infections primarily come from other cattle. Left: time distribution of samples broken down by species. Centre: distribution of pairwise genetic distances. Right: graph representation where genetically similar hosts are clustered. Host pairs classified as being close are highlighted in black.}
	\label{fig:dataS2}
\end{figure}

\begin{figure}
	\includegraphics[width=\textwidth]{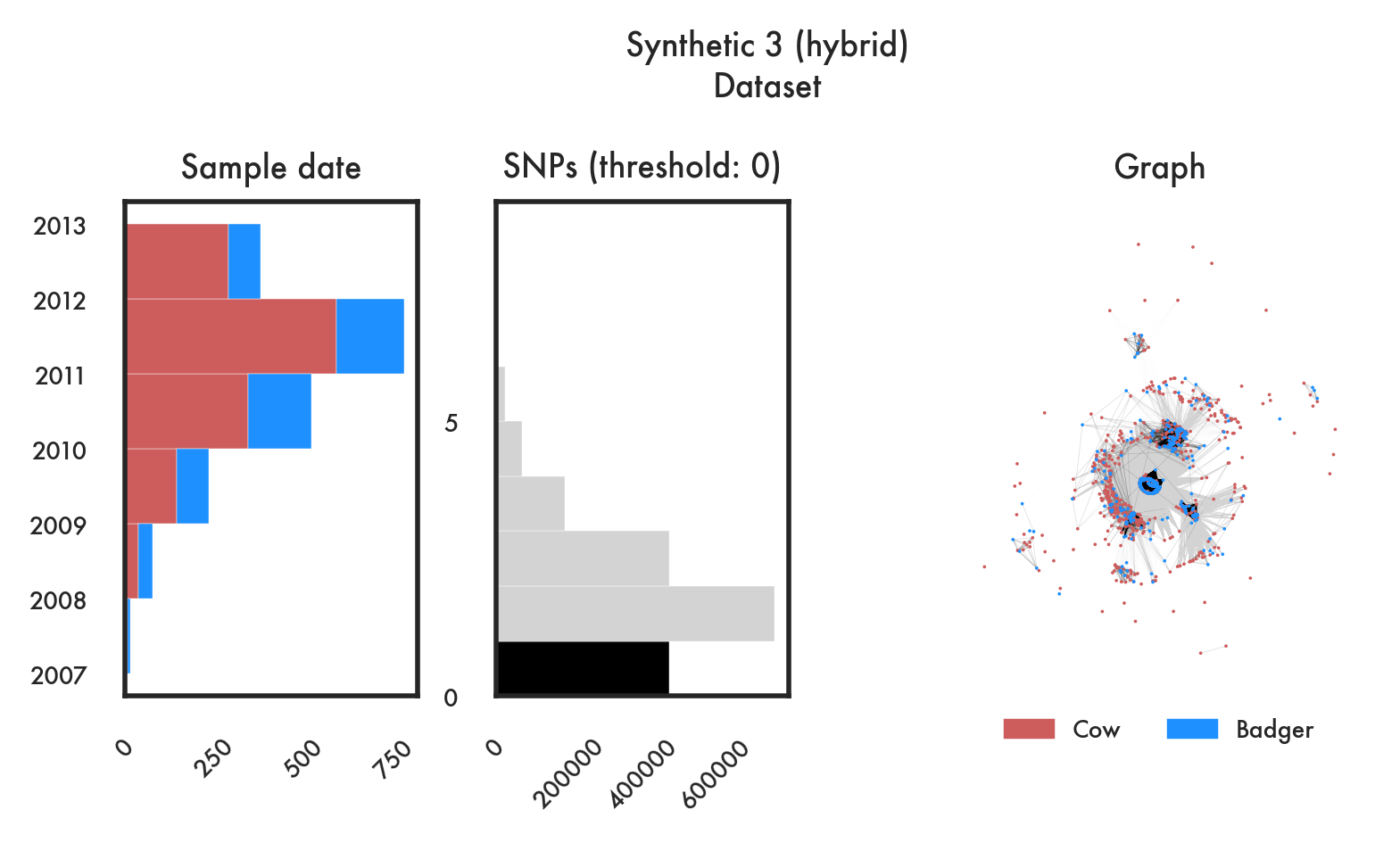}
	\caption{Summary of the Synthetic 3 dataset, run under simulation dynamics where cattle infections primarily come from a mix of other cattle and badgers. Left: time distribution of samples broken down by species. Centre: distribution of pairwise genetic distances. Right: graph representation where genetically similar hosts are clustered. Host pairs classified as being close are highlighted in black.}
	\label{fig:dataS3}
\end{figure}

\begin{table}[h]
	\centering
	{\sffamily
		\footnotesize
		\begin{tabular}{|l|l|}
			\hline
			\textbf{Node attribute} & \textbf{Description} \\
			\hline
			$\texttt{Species}$ & Host species \\
			$\texttt{Date\_Obs}$ & Sample date \\
			$\texttt{Date\_Birth}$ & Known/estimated birth date \\
			$\texttt{Date\_Death}$ & Known/estimated death date \\
			$\texttt{Lon}$ & Longitude \\
			$\texttt{Lat}$ & Latitude \\
			\hline
			\textbf{Edge attribute} & \textbf{Description} \\
			\hline
			$\texttt{N\_Network\_Interactions}$  & (C-C only) Number of direct shared interactions \\
			$\texttt{N\_Network\_Intermediaries}$ & (C-C only) Number of common intermediaries \\
			$\texttt{Physical\_Distance}$ & Euclidean spatial distance \\
			$\texttt{Sample\_t\_Mean}$ & Mean of sample times \\
			$\texttt{Sample\_t\_Diff}$ & Time between sample dates \\
			$\texttt{Death\_t\_Diff}$ & Time between deaths \\
			$\texttt{Birth\_t\_Diff}$ & Time between births \\
			$\texttt{Lifespan\_Overlap}$ & Time overlapping in lifespan \\
			$\texttt{Species\_Type}$ & Species pair type (C-C, B-B, C-B) \\
			$\texttt{Genetic\_Distance}$ & (GNN train set only) pairwise genetic distance \\
			\hline
		\end{tabular}
	}
	\caption{Node and edge attributes in the Synthetic dataset.}
	\label{tab:attributes_synthetic}
\end{table}

\clearpage
\begin{figure}
	\includegraphics[width=\textwidth]{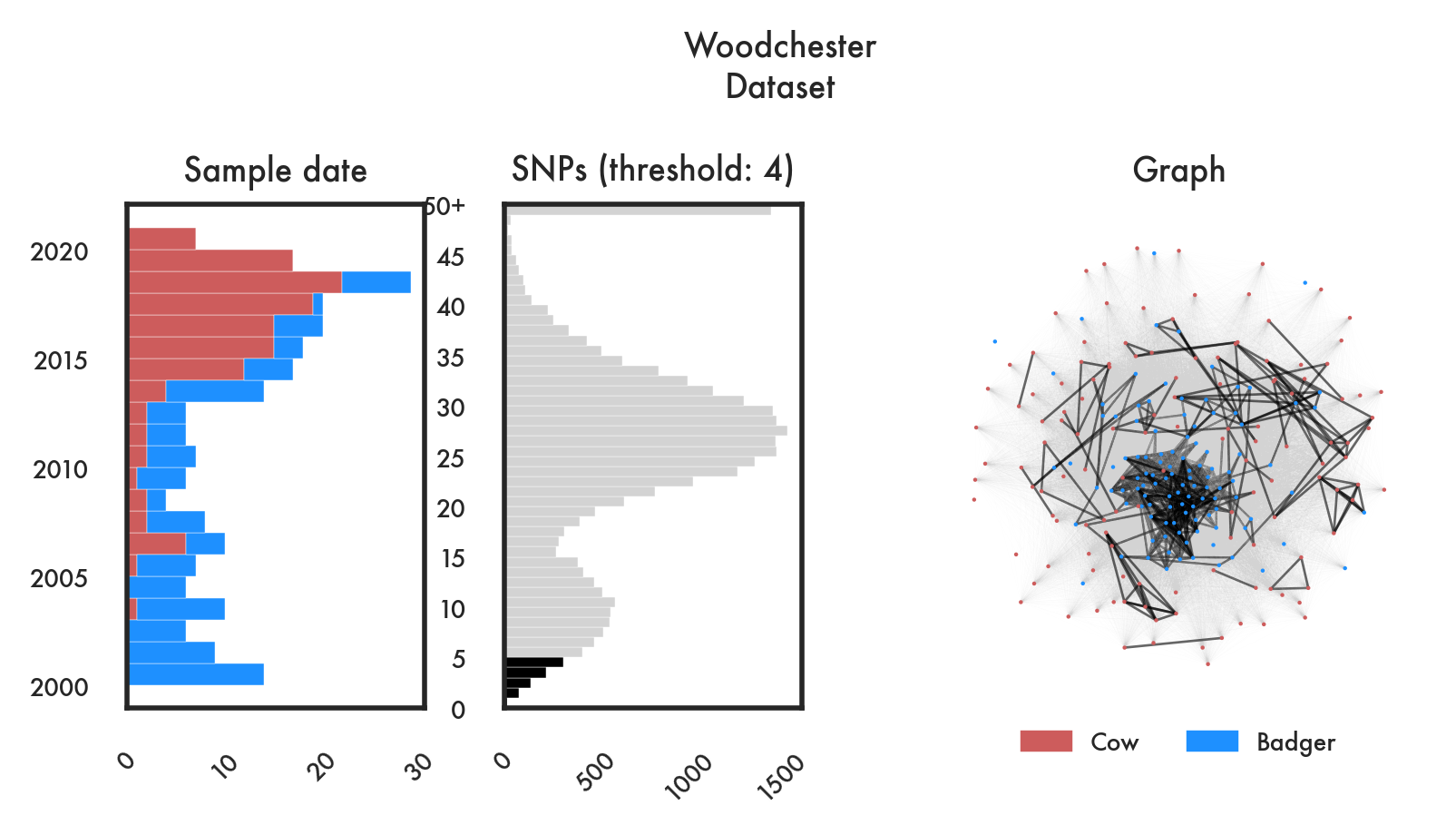}
	\caption{Summary of the Woodchester dataset, from real-world samples taken from the Woodchester Park study. Left: time distribution of samples broken down by species. Centre: distribution of pairwise genetic distances. Right: graph representation where genetically similar hosts are clustered. Host pairs classified as being close are highlighted in black.}
	\label{fig:dataW}
\end{figure}

\begin{table}[ht]
	\centering
	{\sffamily
		\footnotesize
		\begin{tabular}{|l|l|}
			\hline
			\textbf{Node attribute} & \textbf{Description} \\
			\hline
			$\texttt{Species}$ & Host species \\
			$\texttt{Date}$ & Sample date \\
			$\texttt{BirthDate}$ & Known/estimated birth date \\
			$\texttt{DeathDate}$ & Known/estimated death date \\
			$\texttt{xcoord}$ & Spatial coordinate ($x$) \\
			$\texttt{ycoord}$ & Spatial coordinate ($y$) \\
			\hline
			\textbf{Edge attribute} & \textbf{Description} \\
			\hline
			$\texttt{Date\_Diff}$ & Time between samples \\
			$\texttt{Date\_Mean}$ & Mean sampling date \\
			$\texttt{N\_Network\_Interactions}$  & (C-C only) Number of direct shared interactions \\
			$\texttt{N\_Network\_Intermediaries}$ & (C-C only) Number of common intermediaries \\
			$\texttt{Time\_Together}$ & How long hosts spent together (cattle only) \\
			$\texttt{Physical\_Distance}$ & Euclidean spatial distance \\
			$\texttt{Species\_Type}$ & Species pair type (C-C, B-B, C-B) \\
			$\texttt{Genetic\_Distance}$ & (GNN train set only) pairwise genetic distance \\
			\hline
		\end{tabular}
	}
	\caption{Node and edge attributes in the Woodchester dataset.}
	\label{tab:attributes_woodchester}
\end{table}

\clearpage

\begin{figure}
	\includegraphics[width=\textwidth]{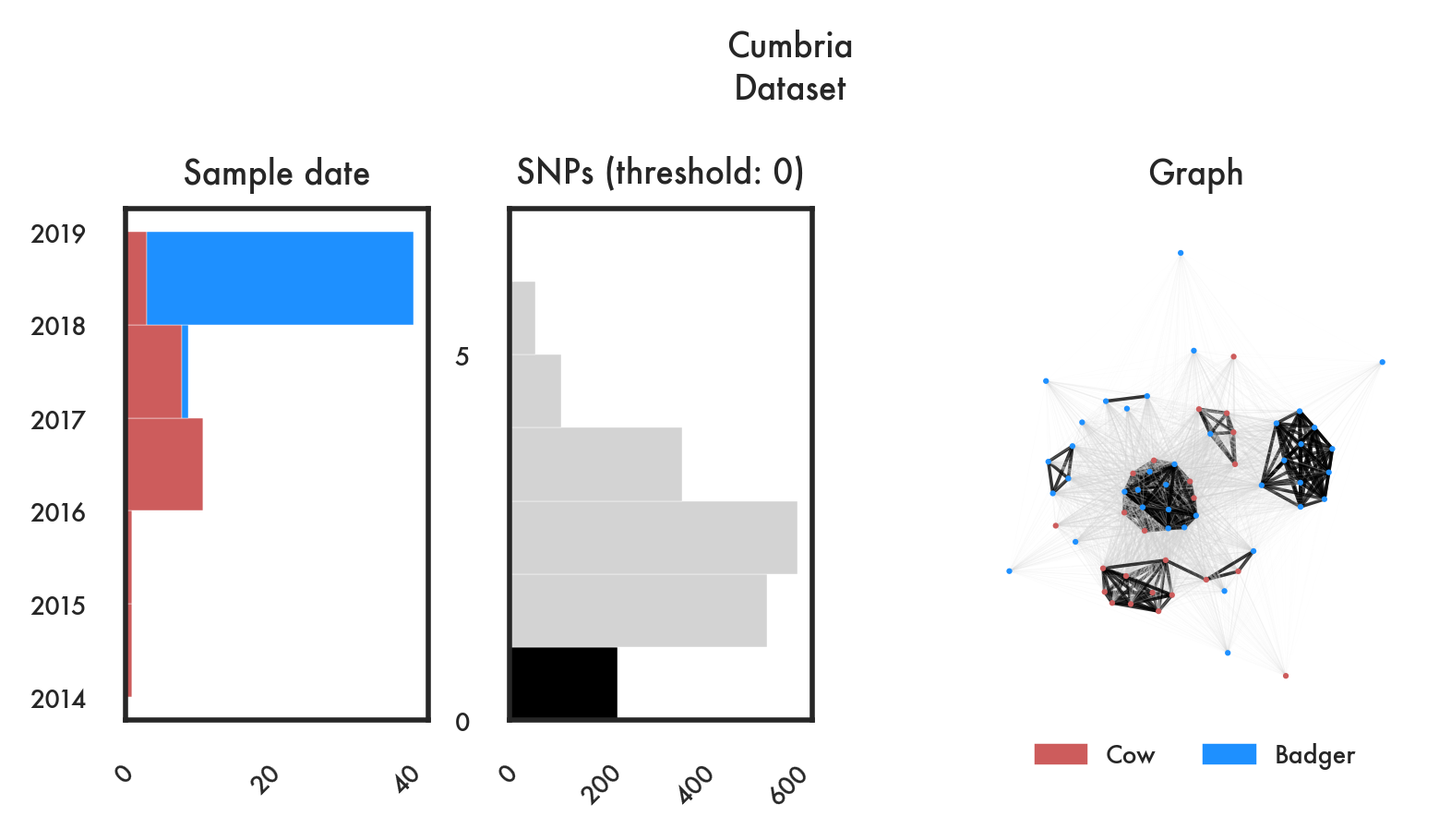}
	\caption{Summary of the Cumbria dataset, from real-world samples taken from the Cumbria bTB outbreak. Left: time distribution of samples broken down by species. Centre: distribution of pairwise genetic distances. Right: graph representation where genetically similar hosts are clustered. Host pairs classified as being close are highlighted in black.}
	\label{fig:dataC}
\end{figure}

\begin{table}[ht]
	\centering
	{\sffamily
		\footnotesize
		\begin{tabular}{|l|l|}
			\hline
			\textbf{Node attribute} & \textbf{Description} \\
			\hline
			$\texttt{Species}$ & Host species \\
			$\texttt{Date}$ & Sample collection date \\
			$\texttt{Sex}$ & Reported sex \\
			$\texttt{BirthDate}$ & Known/estimated birth date \\
			$\texttt{DeathDate}$ & Known/estimated death date \\
			$\texttt{xcoord}$ & Spatial coordinate ($x$) \\
			$\texttt{ycoord}$ & Spatial coordinate ($y$) \\
			\hline
			\textbf{Edge attribute} & \textbf{Description} \\
			\hline
			$\texttt{Date\_Diff}$ & Time between samples \\
			$\texttt{Date\_Mean}$ & Mean sampling date \\
			$\texttt{N\_Network\_Interactions}$  & (C-C only) Number of direct shared interactions \\
			$\texttt{N\_Network\_Intermediaries}$ & (C-C only) Number of common intermediaries \\
			$\texttt{Dist\_Land\_Parcel\_Network}$ & Network distance on land parcel graph \\
			$\texttt{Physical\_Distance}$ & Euclidean spatial distance \\
			$\texttt{Network\_Distance}$ & Distance on contact network \\
			$\texttt{Time\_Together}$ & How long hosts spent together (cattle only) \\
			$\texttt{Species\_Type}$ & Species pair type (C-C, B-B, C-B) \\
			$\texttt{Genetic\_Distance}$ & (GNN train set only) pairwise genetic distance \\
			\hline
		\end{tabular}
		\footnotesize}
	\caption{Node and edge attributes in the Cumbria dataset.}
	\label{tab:attributes_cumbria}
\end{table}

\clearpage
\section{Simulation of bTB outbreaks using Tuberculosis Modelling Initiative simulation model}\label{sec:TBMI}

We use a detailed, individual-based TB model called \emph{Tuberculosis Modelling Initiative} (\emph{TBMI}). In TBMI Cattle may expose other cattle to infection as well as local badgers, who themselves can transmit the disease to one another, or back to local cattle. Cattle populations and movements mirror those in GB from 2006--2020.

The GB mainland is divided into geographical units termed hex cells, each with area of $\sim 45$ sq km and an associated badger population. The model may run over all of GB, but for computational efficiency we isolate 8 hex cells covering the same Cumbria region where real-world samples from the Cumbria dataset were gathered.

\subsubsection*{Model dynamics}
Cattle infections are simulated at an individual holding level under a Susceptible-Exposed-Infectious (SEI) model:
\begin{align}
	\partial_t\, S^i_\C &= - \left(\beta_{\C\to\C}\frac{I^i_\C}{N^i_\C} + \beta_{\B\to\C}I^h_\B\right) S^i_\C \label{eq:cattleSEI} \\
	\partial_t\, E^i_\C &= + \left(\beta_{\C\to\C}\frac{I^i_\C}{N^i_\C} + \beta_{\B\to\C}I^h_\B\right) S^i_\C - \gamma E^i_\C \nonumber  \\
	\partial_t\, I^i_\C &= \gamma E_\C \nonumber 
\end{align}
and badger infections are simulated at a hex-cell level under a Susceptible-Infectious (SI) model:
\begin{align*}
	\partial_t\, S^h_\B &= \mu I^h_\B -\left(\beta_{\C\to \B}\frac{I^h_\C}{N^h_\C} + \beta_{\B\to\B}\left[I^h_\B + \lambda I^{\sim h}_\B\right]\right)S_\B^h \label{eq:badgerSI} \\
	\partial_t\, I^h_\B &= \left(\beta_{\C\to \B}\frac{I^h_\C}{N^h_\C} + \beta_{\B\to\B}\left[I^h_\B + \lambda I^{\sim h}_\B\right]\right)S_\B^h - \mu I^h_\B \nonumber
\end{align*}

Parameter definitions are given in Table~\ref{tab:params}, with the choices used for different synthetic datasets given in Table~\ref{tab:params}. Infections from cattle are frequency-dependent. Infections from badgers are density-dependent. The specific values of the infection rates are not fit to a specific real-world disease scenario, but rather serve to generate host samples over an approximate five-year simulation period, where the underlying transmission tree and transmission dynamics are known exactly.
\begin{table}[h!]
	\centering
	{\sffamily
		\footnotesize
		\begin{tabular}{|l|l|l|l|}
			\hline
			\textbf{Parameter} & \textbf{Description} & \textbf{Units} \\ \hline
			$N_\C^i$ & Cattle in holding $i$ & Number \\ \hline
			$S_\C^i$ & Susceptible cattle in holding $i$ & Number \\ \hline
			$E_\C^i$ & Exposed cattle in holding $i$ & Number \\ \hline
			$I_\C^i$ & Infected cattle in holding $i$ & Number \\ \hline
			$N_\C^i$ & Cattle in all holdings within hex cell $h$ & Number \\ \hline
			$I_\C^i$ & Infected cattle in all holdings within hex cell $h$ & Number \\ \hline
			$S_\B^h$ & Susceptible badgers in hex cell $h$ & Number \\ \hline
			$I_\B^h$ & Infectious badgers in hex cell $h$ & Number \\ \hline
			$I_\B^{\sim h}$ & Infectious badgers in all hex cells adjacent to $h$ & Number \\ \hline
			$\beta_{\C\to\C}$ & (Frequency-dependent) cattle-to-cattle transmission rate & 1/day \\ \hline
			$\beta_{\C\to\B}$ & (Frequency-dependent) cattle-to-badger transmission rate & 1/day \\ \hline
			$\beta_{\B\to\C}$ & (Density-dependent) badger-to-cattle transmission rate & 1/day \\ \hline
			$\beta_{\B\to\B}$ & (Density-dependent) badger-to-badger transmission rate & 1/day \\ \hline
			$\lambda$ & Infection rate multiplier for between-hex-cell badger-badger transmission & Number \\ \hline
			$\mu$ & Birth and death rate of badgers & 1/day \\ \hline
			$\gamma$ & Rate of transition from exposed to infectious (cattle) & 1/day \\ \hline
		\end{tabular}
	}
	\caption{Definitions of parameters included the underlying model equations.}
	\label{tab:params}
\end{table}

\begin{table}[h!]
	\centering
	{\sffamily
		\footnotesize
		\begin{tabular}{|l|c|c|c|c|c|c|}
			\hline
			\textbf{Scenario} & $\beta_{\C\to\C}$ & $\beta_{\B\to\C}$ & $\beta_{\C\to\B}$ & $\beta_{\B\to\B}$ \\
			\hline
			1 (badger-driven) & $1.0\times 10^{-3}$ & $4.0\times 10^{-6}$ & $3.0\times 10^{-4}$ & $6.0\times 10^{-5}$ \\
			\hline
			2 (cattle-driven) & $5.0 \times 10^{-3}$ & $1.0\times 10^{-8}$ & $6.0\times 10^{-4}$ & $5.0\times 10^{-5}$ \\
			\hline
			3 (hybrid) & $2.0\times 10^{-3}$ & $3.0\times 10^{-6}$ & $1.0\times 10^{-3}$ & $5.0\times 10^{-5}$ \\
			\hline
		\end{tabular}
	}
	\caption{Parameter values used in generating the three synthetic datasets. $\beta_{\C\to\C}$ and $\beta_{\C\to\B}$ are frequency-dependent transmission rates. $\beta_{\B\to\B}$ and $\beta_{\B\to\C}$ are density-dependent.}
	\label{tab:beta}
\end{table}

\subsubsection*{Transmission tree construction}
TBMI outputs a list of exposure events. We build a transmission tree by searching for each infection event which animals could have caused that infection. For cattle infectees, this is all infectious cattle in the same holding, and all infectious badgers in the same hex cell. For badger infectees, this is all infectious cattle in the same hex cell, and all infectious badgers in the same hex cell, and neighbouring hex cells. From the set of infectors we choose one with probabilities weighted by the different infection pressures on that host at the time of infection.

\subsubsection*{Modelling pathogen evolution}
For each model trajectory we simulate evolution of the \emph{M. bovis} pathogen. Under this post-hoc pathogen evolution modelling, substitutions of \emph{M. bovis} are neutral. For each infected host we generate a list of times when a substitution on that host's pathogen genome took place (Poisson events with mean rate $\lambda = 0.3$ SNPs / genome / year, beginning from time of exposure). Tracking the substitutions inherited by each new infectee, we assign to each host a pseudo-sequence, with a list substitutions either inherited from its infector or generated between exposure and observation.

\subsubsection*{Sample collection}
At the end of the simulation we record the infection prevalence in cattle, badgers, and also in terms of the numbers of holdings with at least one infection. Finally, we sample infected cattle and associated metadata from these outputs, as one would in a real-world epidemiological investigation. We record from each cow their pseudo-sequence, along with their sample time, sample location, and previous movement history. From each badger we sample each their birth date, death date, and sample location. Animals are sampled at time of death.

\clearpage
\section{Graph neural network model architectures}\label{sec:GNNarchitectures}

\subsection*{Synthetic 1}
\footnotesize{
	\begin{verbatim}
		==========================================================================================
		Layer (type:depth-idx)                   Output Shape              Param #
		==========================================================================================
		GNN                                      [1999000]                 --
		ModuleList: 1-1                        --                        --
		GeneralConv: 2-1                  [2000, 128]               256
		Linear: 3-1                  [2001000, 256]            1,792
		Linear: 3-2                  [2001000, 256]            2,816
		SoftmaxAggregation: 3-3      [2000, 2, 128]            --
		Linear: 3-4                  [2000, 128]               896
		GeneralConv: 2-2                  [2000, 128]               256
		Linear: 3-5                  [2001000, 256]            33,024
		Linear: 3-6                  [2001000, 256]            2,816
		SoftmaxAggregation: 3-7      [2000, 2, 128]            --
		Identity: 3-8                [2000, 128]               --
		GeneralConv: 2-3                  [2000, 128]               128
		Linear: 3-9                  [2001000, 128]            16,512
		Linear: 3-10                 [2001000, 128]            1,408
		SoftmaxAggregation: 3-11     [2000, 1, 128]            --
		Identity: 3-12               [2000, 128]               --
		Sequential: 1-2                        [1999000, 1]              --
		Linear: 2-4                       [1999000, 64]             16,960
		ReLU: 2-5                         [1999000, 64]             --
		Dropout: 2-6                      [1999000, 64]             --
		Linear: 2-7                       [1999000, 8]              520
		ReLU: 2-8                         [1999000, 8]              --
		Dropout: 2-9                      [1999000, 8]              --
		Linear: 2-10                      [1999000, 4]              36
		ReLU: 2-11                        [1999000, 4]              --
		Linear: 2-12                      [1999000, 1]              5
		==========================================================================================
		Total params: 77,425
		Trainable params: 77,425
		Non-trainable params: 0
		Total mult-adds (G): 151.82
		==========================================================================================
		Input size (MB): 111.99
		Forward/backward pass size (MB): 21723.67
		Params size (MB): 0.31
		Estimated Total Size (MB): 21835.97
		==========================================================================================
	\end{verbatim}
}
\subsection*{Synthetic 2}
\footnotesize{
	
	\begin{verbatim}
		
		==========================================================================================
		Layer (type:depth-idx)                   Output Shape              Param #
		==========================================================================================
		GNN                                      [1999000]                 --
		ModuleList: 1-1                        --                        --
		GeneralConv: 2-1                  [2000, 128]               256
		Linear: 3-1                  [2001000, 256]            1,792
		Linear: 3-2                  [2001000, 256]            2,816
		SoftmaxAggregation: 3-3      [2000, 2, 128]            --
		Linear: 3-4                  [2000, 128]               896
		GeneralConv: 2-2                  [2000, 128]               256
		Linear: 3-5                  [2001000, 256]            33,024
		Linear: 3-6                  [2001000, 256]            2,816
		SoftmaxAggregation: 3-7      [2000, 2, 128]            --
		Identity: 3-8                [2000, 128]               --
		GeneralConv: 2-3                  [2000, 128]               128
		Linear: 3-9                  [2001000, 128]            16,512
		Linear: 3-10                 [2001000, 128]            1,408
		SoftmaxAggregation: 3-11     [2000, 1, 128]            --
		Identity: 3-12               [2000, 128]               --
		Sequential: 1-2                        [1999000, 1]              --
		Linear: 2-4                       [1999000, 64]             16,960
		ReLU: 2-5                         [1999000, 64]             --
		Dropout: 2-6                      [1999000, 64]             --
		Linear: 2-7                       [1999000, 8]              520
		ReLU: 2-8                         [1999000, 8]              --
		Dropout: 2-9                      [1999000, 8]              --
		Linear: 2-10                      [1999000, 4]              36
		ReLU: 2-11                        [1999000, 4]              --
		Linear: 2-12                      [1999000, 1]              5
		==========================================================================================
		Total params: 77,425
		Trainable params: 77,425
		Non-trainable params: 0
		Total mult-adds (G): 151.82
		==========================================================================================
		Input size (MB): 111.99
		Forward/backward pass size (MB): 21723.67
		Params size (MB): 0.31
		Estimated Total Size (MB): 21835.97
		==========================================================================================
	\end{verbatim}
}

\subsection*{Synthetic 3}
\footnotesize{
	\begin{verbatim}
		==========================================================================================
		Layer (type:depth-idx)                   Output Shape              Param #
		==========================================================================================
		GNN                                      [1999000]                 --
		ModuleList: 1-1                        --                        --
		GeneralConv: 2-1                  [2000, 128]               256
		Linear: 3-1                  [2001000, 256]            1,792
		Linear: 3-2                  [2001000, 256]            2,816
		SoftmaxAggregation: 3-3      [2000, 2, 128]            --
		Linear: 3-4                  [2000, 128]               896
		GeneralConv: 2-2                  [2000, 128]               256
		Linear: 3-5                  [2001000, 256]            33,024
		Linear: 3-6                  [2001000, 256]            2,816
		SoftmaxAggregation: 3-7      [2000, 2, 128]            --
		Identity: 3-8                [2000, 128]               --
		GeneralConv: 2-3                  [2000, 128]               128
		Linear: 3-9                  [2001000, 128]            16,512
		Linear: 3-10                 [2001000, 128]            1,408
		SoftmaxAggregation: 3-11     [2000, 1, 128]            --
		Identity: 3-12               [2000, 128]               --
		Sequential: 1-2                        [1999000, 1]              --
		Linear: 2-4                       [1999000, 64]             16,960
		ReLU: 2-5                         [1999000, 64]             --
		Dropout: 2-6                      [1999000, 64]             --
		Linear: 2-7                       [1999000, 8]              520
		ReLU: 2-8                         [1999000, 8]              --
		Dropout: 2-9                      [1999000, 8]              --
		Linear: 2-10                      [1999000, 4]              36
		ReLU: 2-11                        [1999000, 4]              --
		Linear: 2-12                      [1999000, 1]              5
		==========================================================================================
		Total params: 77,425
		Trainable params: 77,425
		Non-trainable params: 0
		Total mult-adds (G): 151.82
		==========================================================================================
		Input size (MB): 111.99
		Forward/backward pass size (MB): 21723.67
		Params size (MB): 0.31
		Estimated Total Size (MB): 21835.97
		==========================================================================================
	\end{verbatim}
}

\subsection*{Woodchester}
\footnotesize{
	\begin{verbatim}
		==========================================================================================
		Layer (type:depth-idx)                   Output Shape              Param #
		==========================================================================================
		GNN                                      [28920]                   --
		ModuleList: 1-1                        --                        --
		GeneralConv: 2-1                  [241, 32]                 64
		Linear: 3-1                  [29161, 64]               448
		Linear: 3-2                  [29161, 64]               576
		SoftmaxAggregation: 3-3      [241, 2, 32]              --
		Linear: 3-4                  [241, 32]                 224
		GeneralConv: 2-2                  [241, 32]                 64
		Linear: 3-5                  [29161, 64]               2,112
		Linear: 3-6                  [29161, 64]               576
		SoftmaxAggregation: 3-7      [241, 2, 32]              --
		Identity: 3-8                [241, 32]                 --
		GeneralConv: 2-3                  [241, 32]                 32
		Linear: 3-9                  [29161, 32]               1,056
		Linear: 3-10                 [29161, 32]               288
		SoftmaxAggregation: 3-11     [241, 1, 32]              --
		Identity: 3-12               [241, 32]                 --
		Sequential: 1-2                        [28920, 1]                --
		Linear: 2-4                       [28920, 64]               4,608
		ReLU: 2-5                         [28920, 64]               --
		Dropout: 2-6                      [28920, 64]               --
		Linear: 2-7                       [28920, 8]                520
		ReLU: 2-8                         [28920, 8]                --
		Dropout: 2-9                      [28920, 8]                --
		Linear: 2-10                      [28920, 4]                36
		ReLU: 2-11                        [28920, 4]                --
		Linear: 2-12                      [28920, 1]                5
		==========================================================================================
		Total params: 10,609
		Trainable params: 10,609
		Non-trainable params: 0
		Total mult-adds (M): 296.98
		==========================================================================================
		Input size (MB): 1.39
		Forward/backward pass size (MB): 92.53
		Params size (MB): 0.04
		Estimated Total Size (MB): 93.96
		==========================================================================================
	\end{verbatim}
}

\subsection*{Cumbria}
\footnotesize{
	\begin{verbatim}
		==========================================================================================
		Layer (type:depth-idx)                   Output Shape              Param #
		==========================================================================================
		GNN                                      [1953]                    --
		ModuleList: 1-1                        --                        --
		GeneralConv: 2-1                  [63, 64]                  64
		Linear: 3-1                  [2016, 64]                512
		Linear: 3-2                  [2016, 64]                704
		SoftmaxAggregation: 3-3      [63, 1, 64]               --
		Linear: 3-4                  [63, 64]                  512
		Sequential: 1-2                        [1953, 1]                 --
		Linear: 2-2                       [1953, 64]                8,832
		ReLU: 2-3                         [1953, 64]                --
		Dropout: 2-4                      [1953, 64]                --
		Linear: 2-5                       [1953, 8]                 520
		ReLU: 2-6                         [1953, 8]                 --
		Dropout: 2-7                      [1953, 8]                 --
		Linear: 2-8                       [1953, 4]                 36
		ReLU: 2-9                         [1953, 4]                 --
		Linear: 2-10                      [1953, 1]                 5
		==========================================================================================
		Total params: 11,185
		Trainable params: 11,185
		Non-trainable params: 0
		Total mult-adds (M): 20.83
		==========================================================================================
		Input size (MB): 0.11
		Forward/backward pass size (MB): 3.30
		Params size (MB): 0.04
		Estimated Total Size (MB): 3.46
		==========================================================================================
	\end{verbatim}
}

\end{document}